

\input harvmac
\input epsf
\def\lsim{\mathrel{\rlap{\lower4pt\hbox{\hskip1pt$\sim$}}
    \raise1pt\hbox{$<$}}}         
\noblackbox
\pageno=0\nopagenumbers\tolerance=10000\hfuzz=5pt
\line{\hfill CERN-TH/95-31}
\line{\hfill GeF-TH-2/95}
\vskip 18pt
\centerline{\bf SCALE DEPENDENCE AND SMALL $x$ BEHAVIOUR }
\centerline{\bf OF POLARIZED PARTON DISTRIBUTIONS}
\vskip 36pt\centerline{Richard~D.~Ball\footnote{*}{On leave
 from a Royal Society University Research Fellowship.},
Stefano~Forte\footnote{\dag}{On leave
 from INFN, Sezione di Torino, Italy.}
and Giovanni Ridolfi\footnote{\ddag} {On leave
 from INFN, Sezione di Genova, Italy.}}
\vskip 12pt
\centerline{\it Theory Division, CERN,}
\centerline{\it CH-1211 Gen\`eve 23, Switzerland.}
\vskip 36pt
{\centerline{\bf Abstract}
\medskip\narrower
\ninepoint\baselineskip=9pt plus 2pt minus 1pt
\lineskiplimit=1pt \lineskip=2pt
We discuss perturbative evolution of the polarized structure function $g_1$ in
the $(x,\, Q^2)$ plane, with special regard to the small-$x$ region. We
determine $g_1$ in terms of polarized quark and gluon distributions using
coefficient functions to order $\alpha_s$. At small $x$ $g_1$ then displays
substantial scale dependence, which necessarily implies a corresponding scale
dependence in the large $x$ region. This scale dependence has significant
consequences for the extraction of the first moment from the experimental data,
reducing its value while increasing the error. Conversely, the scale dependence
may  be used to constrain the size of the polarized gluon distribution.
\smallskip}
\vskip 24pt
\line{CERN-TH/95-31\hfill}
\line{February 1995\hfill}

\vfill\eject \footline={\hss\tenrm\folio\hss}

\def\as{\alpha_s}

\def\neath#1#2{\mathrel{\mathop{#1}\limits_{#2}}}

\def\neath#1#2{\mathrel{\mathop{#1}\limits_{#2}}}
\def\gsim{\mathrel{\rlap{\lower4pt\hbox{\hskip1pt$\sim$}}
    \raise1pt\hbox{$>$}}}         
\def\eg{{\it e.g.}}

\def\frac#1#2{{{#1}\over {#2}}}
\def\half{\hbox{${1\over 2}$}}

\def\smallfrac#1#2{\hbox{${{#1}\over {#2}}$}}

\def\GeV{{\rm GeV}}

\catcode`@=11 
\def\slash#1{\mathord{\mathpalette\c@ncel#1}}
 \def\c@ncel#1#2{\ooalign{$\hfil#1\mkern1mu/\hfil$\crcr$#1#2$}}
\def\lsim{\mathrel{\mathpalette\@versim<}}
\def\gsim{\mathrel{\mathpalette\@versim>}}
 \def\@versim#1#2{\lower0.2ex\vbox{\baselineskip\z@skip\lineskip\z@skip
       \lineskiplimit\z@\ialign{$\m@th#1\hfil##$\crcr#2\crcr\sim\crcr}}}
\catcode`@=12 

\def\PR{{\it Phys.~Rev.~}}
\def\PRL{{\it Phys.~Rev.~Lett.~}}
\def\NP{{\it Nucl.~Phys.~}}
\def\NPBPS{{\it Nucl.~Phys.~B (Proc.~Suppl.)~}}
\def\PL{{\it Phys.~Lett.~}}
\def\PRep{{\it Phys.~Rep.~}}

\def\RMP{{\it Rev.~Mod.~Phys.~}}

\def\ZP{{\it Zeit.~Phys.~}}

\def\vol#1{{\bf #1}}\def\vyp#1#2#3{\vol{#1} (#2) #3}

\nref\revs{For recent reviews see G.~Altarelli and G.~Ridolfi,
\NPBPS\vyp{39B}{1995}{106}\semi S.~Forte, preprint CERN-TH.7453/94.}
\nref\SMC{SMC Collaboration,
D.~Adams et al., \PL\vyp{B329}{1994}{399}.}\nref\Eoft{E143
Collaboration, K.~Abe et al., \PRL\vyp{74}{1995}{346}.}
\newsec{Introduction}

Experimental data on the polarized proton structure function $g_1$\revs\ have
substantially improved within the last few months~\refs{\SMC,\Eoft}, and
results of comparable accuracy for the deuteron should be available
soon~\ref\slacpr{E143 Collaboration, R.~Arnold et al., preliminary results
presented at ICHEP94, Glasgow, August 1994.}. Not only are statistical errors
now substantially reduced~\Eoft, but also data which extend to rather small
values of $x$ (of order $0.003$~\SMC) are  available, as well as determinations
of the structure function at different scales for a given value of $x$. This
suggests that it may  be possible to see the effects of perturbative QCD
evolution in the data. Indeed, the contribution of the polarized gluon
distribution to $g_1$ which appears in next-to-leading order, has been put
forward~\ref\alros{G.~Altarelli and G.~G.~Ross, \PL\vyp{B212}{1988}{391}.}  as
one possible way of understanding the observed smallness of the first moment
\eqn\fmom
{\Gamma_1(Q^2)\equiv \int_0^1 g_1(x,Q^2) dx;}
it may now be possible to test for such a contribution by examining the $x$ and
$Q^2$ dependence of the structure function data. Also, the small-$x$ data~\SMC\
indicate (albeit with large experimental uncertainty) a rise of the structure
function at small $x$. Such a rise, even if not present in $g_1$ at low scales,
is generated at higher scales through perturbative QCD
evolution~\ref\ahros{M.~A.~Ahmed and G.~G.~Ross, \PL\vyp{B56}{1975}{385}.} by a
similar mechanism to that which drives the rise in the unpolarized structure
function $F_2^P$ at small $x$~\ref\deruj{A.~De~Rujula, S.L.~Glashow,
H.D.~Politzer,
S.B.~Treiman, F.~Wilczek and A.~Zee, \PR\vyp{D10}{1974}{1649}.}
recently observed at HERA~\ref\das {R.~D.~Ball and
S.~Forte, \PL\vyp{B335}{1994}{77}; \PL\vyp{B336}{1994}{77}.}.

A quantitative understanding of the scale dependence and the small-$x$
behaviour of the structure function is required in order to extract accurately
its  moments  from the data: because data are taken at different values of
$Q^2$ for each $x$ bin, and over a limited range in $x$, it is necessary to
evolve them to a common scale $t=\ln(Q^2/\Lambda^2)$ and extrapolate them to
all $0\le x\le1$ in order to determine any moment of $g_1$. The evolution is
usually performed by the experimental collaborations~\refs{\SMC,\Eoft} by
assuming~\ref\ellkar{J.~Ellis and M.~Karliner, \PL\vyp{B313}{1993}{131}.}
that the virtual photon scattering asymmetry\foot{We follow the
notations and conventions of ref.~\revs, to which the reader is referred for
theoretical background.}
\eqn\asym
{A_1(x,Q^2)\equiv{ \sigma_{1/2}- \sigma_{3/2}\over
\sigma_{1/2}+ \sigma_{3/2}}=g_1(x,Q^2)
\frac{2x\left[1+R(x,Q^2)\right]}{F_2(x, Q^2)}}
is scale independent, so that the scale dependence of $g_1$ is given by that of
the unpolarized structure function $F_2$, which is known rather accurately.
This can however be a rather poor approximation: as we shall see below, at
small $x$ ${d\ln A_1\over dt}$ diverges, and thus a large error can be made if
it is assumed that $A_1$ is scale independent in that region. Likewise, the
small-$x$ extrapolation is usually made by assuming Regge behaviour, which is
not necessarily preserved by  perturbative QCD evolution. This is especially
dangerous since the contribution of the unmeasured small-$x$ region to the
first moment $\Gamma_1$ eq.~\fmom\ is extrapolated from the lowest $x$ data,
which are necessarily taken at rather low values of $Q^2$. The small-$x$
contribution accounts for about $5\%$ of the first moment \Eoft, and because of
large cancellations can give some $20$ or $30\%$ of the physically interesting
singlet component of $\Gamma_1$\revs.

A detailed investigation of the expected $x$ and $Q^2$ dependence of $g_1$,
especially at small $x$, is thus called for. A first study of the scale
dependence of the asymmetry eq.~\asym\ was performed in
ref.~\ref\ANR{G.~Altarelli, P.~Nason and G.~Ridolfi,
\PL\vyp{B320}{1994}{152}.}, where the one-loop Altarelli-Parisi evolution
equations were linearized in $\ln t$ to determine the approximate correction to
be applied to the asymmetry data.  The  corresponding correction to the
measured first moment turns out to be very small (of the order of a few
percent, i.e. smaller than statistical uncertainties). However, the effect on
individual data points may be appreciable; also, since these corrections are
correlated, relatively small effects on single data points can lead to a
significant change of the integral computed over a finite region of $x$. Of
course, these effects  would approximately cancel out if the integral could be
measured over the whole range $0\le x\le 1$ (due to the one-loop conservation
of the first moment of $g_1$). Such scale dependence corrections are thus
increasingly significant as both the statistical quality and coverage of the
small-$x$ region of the data improves. It is then important to investigate the
effects of the next-to-leading order coupling of polarized gluons to $g_1$, in
particular  at small $x$, where the polarized gluon distribution will diverge.

In the present paper we will analyze the $x$ and $Q^2$ dependence of $g_1$,
including all the available theoretical~\ref\willy{ E.B.~Zijlstra and
W.L.~van~Neerven, \NP\vyp{B417}{1994}{61}.} and experimental~\refs{\SMC,\Eoft}
information. In particular, we will study the effects of including the coupling
(which only appears at two loops) of the polarized gluon distribution to $g_1$,
and discuss the perturbative QCD prediction for the behaviour of $g_1$ at small
$x$. Even though a fully consistent two-loop determination of $g_1$ is not yet
possible due to the lack of knowledge of the full set of two-loop
Altarelli-Parisi splitting functions, we will  show that an approximate
treatment suggests that evolution effects are significantly enhanced by the
gluon contribution to $g_1$, and that the inclusion of this contribution is
necessary to understand current data.  The size of these effects is essentially
controlled by the size of the polarized gluon distribution; because this is
unknown, it introduces an uncertainty in the determination of $g_1$. However,
especially  in the small $x$ region, interference  between the perturbative
evolution of gluon and quark contributions may lead to large fluctuations of
$g_1$, so that the data may be used to constrain the size of this gluon
contribution.  We will discuss how present and future data can be used to
determine the nature of the small-$x$ behaviour of polarized parton
distributions, thereby extracting information on the polarized gluon
distribution. Also, we will discuss the impact of these evolution effects on
the extractions from the data of the first moment of $\Gamma_1$, and show it to
be not negligible at the level of accuracy of current experiments.

\newsec{Scale Dependence}

Our starting point is the relation between the polarized structure function
$g_1$ and the polarized quark and gluon distributions $\Delta q_i$ and $\Delta
g$, which in general reads\nref\guidorev{G.~Altarelli,
\PRep\vyp{81}{1982}{1}.}~\refs{\willy,\guidorev}
\eqn\gone
{\eqalign{g_1(x,t)=
\frac{\langle e^2\rangle}{2}\int_x^1\! \frac{dy}{y}\,\Big\{
C_q^{\rm S}(\smallfrac{x}{y},\as(t))\Delta\Sigma(y,t)
      &+ C_q^{\rm NS}(\smallfrac{x}{y},\as(t))\Delta q^{\rm NS}(y,t)\cr
      &+ 2n_f C_g(\smallfrac{x}{y},\as(t))\Delta g(y,t)\Big\},}}
where $\Delta \Sigma$ and $\Delta q^{\rm NS}$ are respectively the
singlet and nonsinglet quark distributions
\eqn\qsing
{\Delta\Sigma(x,t)= \sum_{i=1}^{n_f}
(\Delta q_i(x,t)+\Delta\bar q_i(x,t)),\qquad
\Delta q^{\rm NS}(x,t)=\sum_{i=1}^{n_f}
\frac{e^2_i-\langle e^2\rangle}{\langle e^2\rangle}
(\Delta q_i(x,t)+\Delta\bar q_i(x,t)),}
and $n_f$ is the number of active flavours, with electric charge
$e_i$, $\langle e^2\rangle=\sum e_i^2/n_f$.
The perturbative part of the $x$ and $t$ dependence of the polarized quark and
gluon distributions is given by the Altarelli-Parisi equations~\ref\ap{
G.~Altarelli and G.~Parisi, \NP\vyp{B126}{1977}{298}.}: the singlet quark and
the gluon mix according to
\eqn\aps{\eqalign{
\frac{d}{dt}\Delta\Sigma(x,t)
&=\frac{\as(t)}{2\pi}\int_x^1\! \frac{dy}{y}\,\left[
P_{qq}^{\rm S}(\smallfrac{x}{y},\as(t))\Delta\Sigma(y,t)
          + 2n_fP_{qg}(\smallfrac{x}{y},\as(t))\Delta g(y,t)\right]\cr
\frac{d}{dt}\Delta g(x,t)
&=\frac{\as(t)}{2\pi}\int_x^1\! \frac{dy}{y}\,\left[
P_{gq}(\smallfrac{x}{y},\as(t))\Delta\Sigma(y,t)
                 + P_{gg}(\smallfrac{x}{y},\as(t))\Delta g(y,t)\right],}}
while the nonsinglet quark evolves independently as
\eqn\apqns
{\frac{d}{dt}\Delta q^{\rm NS}(x,t)
=\frac{\as(t)}{2\pi}\int_x^1\! \frac{dy}{y}\,
P_{qq}^{\rm NS}(\smallfrac{x}{y},\as(t))\Delta q^{\rm NS}(y,t).}
In eqs.~\gone-\qsing\ we have taken both the factorization
scale and the renormalization scale equal to $Q^2$, so that all
scale dependence appears through $t=\ln\frac{Q^2}{\Lambda^2}$.
A different choice of factorization scale is discussed
in ref.~\willy.

The coefficient functions which relate the polarized parton distributions to
$g_1$ according to eq.~\gone\ are given in the leading logarithmic
approximation (used in ref.~\ANR) by
\eqn\llogc
{C_q^{(0),\,{\rm S}}(x,\as)=C_q^{(0),\,{\rm NS}}(x,\as)=\delta(1-x),
\qquad C_g(x,\as)=0,}
so that $g_1$ is just a linear combination of polarized quark
distributions, whose $Q^2$ dependence is
entirely specified by eqs.~\aps-\apqns\ (with the leading order
splitting functions calculated in ref.~\ap).

Beyond leading order, splitting functions and coefficient functions are no
longer universal, hence even though the scale dependence of the (observable)
structure function $g_1$ is determined uniquely, at least up to higher order
corrections, its separation into contributions due to quarks and gluons is
scheme dependent and thus essentially arbitrary. This ambiguity is somewhat
constrained in the polarized case because of the anomalous conservation law
satisfied by the singlet axial current~\ref\roman{See e.g. R.~Jackiw in
S.~Treiman, R.~Jackiw, B.~Zumino and E.~Witten, {\it Current Algebra and
Anomalies} (World Scientific, Singapore, 1985).}, whose nucleon matrix element
is proportional to the singlet component of the first moment $\Gamma_1$
eq.~\fmom\ of $g_1$~\revs. The anomaly has the important implication \alros\
that the gluon contribution to $g_1$, although formally of order $\alpha_s$,
does not decouple as $Q^2\to\infty$. We thus expect  the leading order
approximation to $g_1$ to be particularly poor in this case.

Unfortunately, while all the coefficient functions have been calculated up to
order $\alpha_s^2$ in ref.~\willy, only the $O(\alpha_s)$ corrections to the
splitting functions $P^{qg}$ and $P^{qq}$ are known at present. Therefore we
cannot yet perform a fully consistent next-to-leading order analysis. We will
thus solve the Altarelli-Parisi eqs.~\aps~\apqns\ at one loop, but include
coefficient functions through order $\alpha_s$. The gluon coefficient function
would be specified uniquely only once the next-to-leading contributions to the
splitting functions are known in a particular scheme. We can partially fix it
by insisting that the first moment of the polarized quark distribution be
conserved by perturbative evolution. The first moment of the gluon coefficient
function is then fixed by the anomaly equation to be~\revs \eqn\glucoup
{\int_0^1 C_g(x,\alpha_s) dx=-\frac{\as}{4\pi}.} With this choice, the gluon
density coincides with that which can be measured in independent hard
processes~\ref\ccm{R.~D.~Carlitz, J.~C.~Collins and A.~H.~Mueller,
\PL\vyp{B214}{1988}{229}.}, in that all soft contributions to $g_1$ are
absorbed in the parton distributions and do not contribute to coefficient
functions~\ref\polfac{G.~Altarelli and B.~Lampe, \ZP\vyp{C47}{1990}{315}\semi
W.~Vogelsang, \ZP\vyp{C50}{1991}{275}.}. Furthermore, the first moment of
$\Delta \Sigma$ coincides then with the nucleon matrix element of the canonical
quark helicity operator~\ref\sfpol{S.~Forte, \NP\vyp{B331}{1990}{1}.}. We will
further choose the    higher moments of $C_g$ to be equal to the
first~\ref\astir{G.~Altarelli and W.J.~Stirling,
{\it Particle~World~}\vyp{1}{1989}{40}.}. We may then test the sensitivity of
our results to the inclusion
of higher order corrections by choosing alternative forms of the gluon
coefficient function which still satisfy eq.~\glucoup, but result from
different regularizations of infrared singularities~\refs{\alros,\polfac}. In
this respect our approach is the same as that adopted in
ref.~\ref\stirge{T.~Gehrmann and W.J.~Stirling, Durham preprint DTP/94/38
(1994).}. As we shall see, evidence from the data supports the expectation that
including the gluon coupling to $g_1$ by enforcing eq.~\glucoup\ improves the
quality of the leading order calculation, even though of course only a
consistent two loop treatment  will allow a definite conclusion. The inclusion
of $O(\alpha_s)$ corrections to the quark coefficient functions is
known\nref\ellkarbis{J.~Ellis and M.~Karliner,
\PL\vyp{B341}{1995}{397}}~\refs{\ellkarbis,\revs} to have
a significant effect on the determination of $\Gamma_1$ eq.~\fmom.

With this choice, the Mellin  transforms of the coefficient functions which we
will use to determine $g_1$, defined according to $C(N,\alpha_s)=
\int_0^1\!dx\, x^{N-1} C(x,\alpha_s)$ are
\eqn\melcf
{\eqalign{C_q^{\rm NS}(N,\as)=1
+&{\alpha_s\over 4\pi}C_F\Bigg\{
\left(\psi(N)+\gamma\right)\left[3+2\left(\psi(N+2)+\gamma\right)\right]\cr
         +&2\left[\psi^\prime(N)-{\pi^2\over6}\right]-9+{6\over N}\Bigg\}
+O(\as^2)\cr
C_q^{\rm S}(N,\as)&=C_q^{\rm NS}(N,\as)+O(\as^2)\cr
C_g(N,\as)&=-{\alpha_s\over 4\pi}+O(\as^2),\cr}}
where $\psi(N)=d\log\Gamma(N)/dN$.
The explicit relation between the matrix element of the current and
the first moments of the quark and gluon distributions is then
\eqn\axfmom
{\eqalign{&\langle p,s|j^\mu_5|p,s\rangle
=M s^\mu a_0(Q^2),\cr
& a_0(Q^2)=\int_0^1\!dx\,\left[\Delta \Sigma(x,Q^2)-n_f{\alpha_s\over 2\pi}
\Delta g(x,Q^2)\right]\cr}}
where $p$, $M$ and $s$ are the momentum, mass and spin of the nucleon. The
(scale dependent) coefficient of proportionality between $a_0$ and the singlet
component of $\Gamma_1$ can  be determined in terms of the first moments of the
quark and gluon coefficient functions comparing eq.~\axfmom\ with eq.~\gone.

The anomalous dimensions analogously obtained by
Mellin transform of the Altarelli-Parisi splitting functions in eq.~\aps\
are
\eqn\melap
{\eqalign{\gamma_N^{qq}&=C_F\left[{3\over2}+{1\over
N(N+1)}-2\left(\psi(N+1)+\gamma\right) \right]+O(\as)\cr
\gamma_N^{qg}&=T_F\frac{N-1}{N(N+1)}+O(\as)\cr
\gamma_N^{gq}&=C_F\frac{N+2}{N(N+1)}+O(\as)\cr
\gamma_N^{gg}&=C_A\left[{11\over6}-{n_f\over 9}+\frac{4}{N(N+1)}-
2\left(\psi(N+1)+\gamma\right)\right]
+O(\as),\cr}}
where, for SU(3) color, $C_F={4\over3}$, $C_A=3$ and $T_F={1\over2}$.

\newsec{Small-$x$ Behaviour}

We can now discuss the QCD predictions for the small-$x$ behaviour of $\Delta
q$, $\Delta g$, and thus $g_1$. The treatment closely follows that of the
unpolarized case~\das. At small $x$, the Altarelli-Parisi evolution is
dominated by the lowest moments of the splitting function~\deruj; since these
are singular at small enough $N$ the leading behaviour is found by expanding
the anomalous dimensions eq.~\melap\ in powers of $N$ around the rightmost
singularity~\ref\ein{ M.~B.~Einhorn and J.~Soffer, \NP\vyp{B74}{1986}{714}.}:
\eqn\sxap
{\eqalign{\gamma_N^{qq}&=C_F\left[{1\over N}+{1\over2}+O(N) \right]+O(\as)\cr
\gamma_N^{qg}&=T_F\left[-{1\over N}+2+O(N) \right]+O(\as)\cr
\gamma_N^{gq}&=C_F\left[{2\over N}-1+O(N) \right]+O(\as)\cr
\gamma_N^{gg}&=C_A\left[{4\over N}-\left({n_f\over9}+{13\over6}\right)+
O(N) \right]
+O(\as).\cr}}
Following the procedure described in ref.~\das, i.e. diagonalizing the matrix
of anomalous dimensions, inverting the Mellin transform, substituting the
resulting splitting functions into the evolution equation \aps, and
differentiating with respect to $x$, the eigenvectors $v^\pm=(\Delta
q^\pm,\Delta g^\pm)$ are found to evolve in the $(x,t)$ plane according to the
pair of wave-like equations
\eqn\weq{\frac{\partial^2}{\partial\xi\partial\zeta}
v^\pm(\xi,\zeta)+\delta_\pm\frac{\partial}{\partial\xi}
v^\pm(\xi,\zeta)=\gamma_\pm^2
v^\pm(\xi,\zeta)}
The relevant variables are
\eqn\xizeta
{\xi =\ln\smallfrac{x_0}{x},\qquad
\zeta =\ln\smallfrac{t}{t_0},}
and $x_0$ and $t_0\equiv\ln(Q_0^2/\Lambda^2)$ are arbitrary reference values of
$x$ and $t$, chosen in such a way that the approximate form of the splitting
functions eq.\sxap\ is applicable for $x\lsim x_0$ and $t\gsim t_0$. The
coefficients $\gamma_\pm$ and $\delta_\pm$ are determined by the
diagonalization of the singlet anomalous dimension matrix \sxap: the
eigenvalues are $\lambda_\pm=\half\beta_0\big(\gamma^2_\pm/N + \delta^\pm
+O(N)\big)$, with
\eqn\sxcoefs
{\eqalign{\gamma_\pm^2&={8\over 33-2 n_f}\left(5\pm4\sqrt{
1-{3n_f\over 32}}\right)\cr
\delta_\pm&=\frac{35+2n_f\pm
43\left({1-11n_f/86\over\sqrt{1-3n_f/32}}\right)}{2\left(33-2n_f\right),}\cr}}
and $\beta_0\equiv 11-\smallfrac{2}{3}n_f$ is the leading coefficient of
the QCD beta function. The eigenvectors $v^\pm=(\Delta q^\pm,\Delta g^\pm)$ are
given (to leading order in $N$) by
\eqn\evecs
{\Delta q^\pm= -2\left(1\mp\sqrt{1-{3n_f\over 32}}\right)\Delta g^\pm.}
Notice that the eigenvectors are not orthogonal.

The small-$x$ evolution equations satisfied by $v^\pm$ are thus of exactly the
same form as those satisfied by $G(x,Q^2)=x g(x,Q^2)$, where $g$ is the
unpolarized gluon distribution~\das. There are two basic differences between
the polarized and the unpolarized case. Firstly, since the polarized anomalous
dimensions have their rightmost singularity at $N=0$, while the unpolarized
ones have it at $N=1$,  parton distributions here play the same role as parton
distributions multiplied by $x$ in the unpolarized case. Furthermore, since in
the polarized case each entry in the matrix of anomalous dimensions has a
simple pole at $N=0$, the polarized quark and gluon distributions evolve
together on the same footing and the eigenvector of evolution is a linear
combination of them, while in the unpolarized case, since only $\gamma^{gq}$
and $\gamma^{gg}$ are singular, the gluon distribution evolves independently
according to a wave equation of the form \weq, and the quark is then determined
by it.

These differences apart, the unpolarized and polarized cases have much in
common since the form of the equation\weq\ is the same. In particular:
\item{(i)} The equations \weq\ are linear and causal. Their solution at
$(x^\prime,t^\prime)$ is determined by the knowledge of boundary conditions
given on the $x=x_0$ and $t=t_0$ axes for all $x^\prime\le x\le x_0$ and
$t^\prime\ge t\ge t_0$. \item{(ii)} The equations are symmetric in $\xi$ and
$\zeta$, modulo the damping term proportional to $\delta_\pm$. Their solutions
will  display this symmetry, unless $\xi \ll\zeta$, in which case the damping
term is increasingly important because the small-$N$ expansion eq.~\melap\ is
beginning to break down; any further asymmetry will be due to the boundary
conditions. \item{(iii)} The general solution can be determined exactly in
terms of the Green's function $I_0(2\gamma_\pm\sqrt{\xi\zeta})$ of the wave
operator:
\eqn\soln{\eqalign{v^{\pm}(\xi,\zeta)=
            I_0&\big(2\gamma_\pm\sqrt{\xi\zeta}\big)
e^{-\delta_\pm\zeta}v^\pm(0,0)
            +\int_0^\xi d\xi' I_0\big(2\gamma_\pm\sqrt{(\xi-\xi')\zeta}\big)
            e^{-\delta_\pm\zeta}\frac{\partial}{\partial\xi'}v^\pm(\xi',0)\cr
           &+\int_0^\zeta d\zeta' I_0\big(2\gamma_\pm\sqrt{
                              \xi(\zeta-\zeta')}\big)
                                   e^{\delta_\pm(\zeta'-\zeta)}
 \Big(\frac{\partial}{\partial\zeta'}v^\pm(0,\zeta')+\delta_\pm
v^\pm(0,\zeta')\Big).\cr}}
\item{(iv)} The asymptotic form of the solutions fall into two classes,
according to the behaviour of the boundary conditions. Let us consider for
definiteness the boundary conditions along the $\zeta=0$ axis, i.e., as a
function of $x$ at fixed $t=t_0$. For boundary conditions on both $\Delta q$
and $\Delta g$ which are less singular than any power of $x$, the asymptotic
behaviour for large $\xi$ and $\zeta$ will be universal and given by the
asymptotic form of the Bessel function $I_0$:
\eqn\asuniv{v^\pm(\xi,\zeta)\sim{1\over\sqrt{4\pi\gamma_\pm\sqrt{\xi\zeta}}}
 \exp\left\{2\gamma_\pm\sqrt{\xi\zeta}-\delta_\pm\zeta\right\}.}
This behaviour holds as $\xi\zeta\to\infty$ along any curve such that
${\xi\over\zeta}\to\infty$.
If instead the boundary conditions have a
power-like singularity of the form $x^{-\lambda}$, the asymptotic
result eq.~\asuniv\ only applies to the
region $\xi< \zeta {\gamma\over\lambda}$, whereas for larger values of $\xi$
we have  asymptotically
\eqn\ashard
{v^\pm(\xi,\zeta)\sim
\exp\left\{\lambda\xi+
\left(\smallfrac{\gamma_\pm^2}{\lambda}-\delta_\pm\right)\zeta\right\},}
i.e., the leading asymptotic behaviour reproduces the boundary
condition. If the boundary conditions on $\Delta q$ and $\Delta g$ are
different, but one (or both) has a power-like singularity, the most
singular behaviour will always dominate asymptotically because of the
mixing introduced by the evolution.

The properties of the small-$x$ solution to the evolution equations allow
us to make some qualitative predictions on the expected behaviour of $g_1$ at
small $x$. First, $|g_1|$ is expected generically to rise at small
$x$~\ahros, at least as fast as eq.~\asuniv, i.e. stronger than any power
of $\ln {1\over x}$. This suggests that any rise of
the boundary condition softer than this
will soon be masked by the perturbative
rise and thus not directly observable. Although both
$\gamma_+$ and $\gamma_-$ are positive, $\gamma_+$ is larger so
asymptotically the eigenvector $v^+$ eq.~\evecs\ will dominate, and,
at small $x$, $\Delta q$ and $\Delta g$ will eventually have the opposite
sign. The sign of the rise of $g_1$ is however not a priori determined,
but rather is sensitive to the relative importance of the quark and
gluon polarizations on the boundaries:
if $\Delta g $ is sufficiently large and positive, then asymptotically $g_1$
will tend to minus infinity. The onset of the asymptotic behaviour will
lead, for
some boundary
conditions, to interesting interference effects.
The shape of the rise will be determined by the qualitative form of
the boundary conditions: it will take the universal form eq.~\asuniv\ for
sufficiently soft boundary conditions, while it will tend to reproduce the
rise in the boundary condition itself for harder ones.
This is displayed in fig.~1, where we show
a contour plot of the leading small-$x$
form of $g_1$ computed by assuming that the Altarelli-Parisi equations
reduce to their small-$x$ form eq.~\weq, with
``soft'' (fig.~1a)  and ``hard'' (fig.~1b) boundary conditions.
If the boundary conditions are soft it is also
possible to determine the asymptotic behaviour of the experimentally measured
asymmetry eq.~\asym: given that the unpolarized structure function measured
at HERA already displays the corresponding universal asymptotic form~\das,
we have
\eqn\asyas
{A_1(x,Q^2)\sim
x\exp\left[(\sqrt{2}-1)2\gamma\sqrt{\xi\zeta}
-\smallfrac{2}{11}\delta\zeta\right],}
where $\gamma$ and $\delta$ are as given in ref.~\das, and we have neglected
some very small $n_f$ dependent corrections. Thus, even though at small $x$
the asymmetry falls as $x$, $\ln(\smallfrac{1}{x}A_1)$ grows linearly
with $\sqrt{\ln(1/x)\ln t}$.

Putting everything together,
we expect that if sufficiently numerous and precise data  on $g_1$ at small
$x$ and large $Q^2$ are available (say, $x<0.1$ and $Q^2> 10$ GeV$^2$) then
it should be possible, just as in the unpolarized case \das, to
determine experimentally whether the small-$x$
structure function and parton distributions are soft or hard. Clearly,
this is not yet the case, since only a couple of the SMC data points~\SMC\ and
none of the E143 data \Eoft\ fall within this range. This is demonstrated
explicitly in fig.~1, where a scatter plot of the small $x$ data
is superimposed on the
leading predicted asymptotic small $x$ behaviour for soft
and hard boundary conditions: it is clear that
the data are insufficient to distinguish between the two
scenarios simply because only a few of the experimental points fall
close to the asymptotic small-$x$ region.

We expect however
that even with a few data at small $x$ and reasonable $Q^2$ it should
be possible to constrain the size  of the gluon polarization, given the
very strong sensitivity of $g_1$ and its scale dependence to
it. Finally, because the small $x$ eigenvector is
a linear combination of quark and gluon (with a large coefficient
multiplying the
gluon contribution), and because both the quark and the gluon distributions
grow at small $x$ but with opposite signs, we expect in general the
appearance in $g_1$ of fluctuations due to interference between
quark and gluon contributions, until the asymptotic behaviour sets
in. The observation of such fluctuations would put a strong constraint
on the shape and size of the gluon,
although presumably now in a non-universal way (i.e., dependent on the
particular form of the boundary conditions).

\newsec{The Data}

In order to establish some more quantitative results, we
need direct input from the experimental
data. We use the SMC and E143
determinations of the scattering asymmetry
eq.~\asym~\refs{\SMC,\Eoft}, together with the NMC determination of the
unpolarized structure function $F_2$~\ref\NMC{NMC
Collaboration, P.~Amaudruz et al., \PL\vyp{B295}{1992}{159}.} and the
SLAC determination~\ref\Rslac{L.~W.~Whitlow et al.,
\PL\vyp{B250}{1990}{193}.} of the ratio $R(x,Q^2)$ of the
longitudinal to transverse virtual photoabsorption cross section, to obtain
values for $g_1(x,Q^2)$ according to eq.\asym\foot{For consistency we
do not include kinematic higher
twist corrections in determining $g_1$ from $A_1$, as was done in
ref.~\Eoft, since these are not included in the extraction~\NMC\
of $F_2$ from the data, and because there exists no systematic treatment of
higher twist corrections to either $g_1$ or $F_2$.}.
This provides us with a determination of $g_1$ along a set of points
$(x_k,Q_k^2)$ in the $(x,Q^2)$
plane; due to the kinematic coverage of the present experiments these points
lie for each experiment on a  curve which is
a monotonically rising function $Q_{\rm exp}^2(x)$.
We then choose a physically motivated parametrization of the initial polarized
quark and gluon distributions $\Delta q_i$ and $\Delta g$ as a function of $x$
for all $0\le x \le 1$ at a
reference starting scale $Q^2=Q_0^2$; $g_1$ is then determined for
all $Q^2\ge Q_0^2$, and in particular at each data point
$(x_k,Q_k^2)$, by means of eq.~\gone\ and the evolution
equations~\aps,\apqns. The free parameters in the initial
parametrization are  determined by a best fit of the
evolved distribution to all the available data.
Notice that, because the Altarelli-Parisi evolution is causal, the structure
function determined in this way for all points $(x,Q^2)$ such that
$Q^2>Q_{\rm exp}^2(x)$ is independent of the specific parametrization, to
the extent that the data determine the shape of $g_1$ along the line
$Q_{\rm exp}^2(x)$. The only real model
dependence here comes in the way $g_1$ is decomposed into quark singlet
and nonsinglet and gluon contributions.

We parametrize the initial quark and gluon distributions as
\eqn\parm
{\eqalign{\Delta\Sigma(x,Q_0^2)&=N\left(\alpha_q,
\beta_q,a_q\right)
\eta_qx^{\alpha_q}
(1-x)^{\beta_q}(1+a_qx)\cr
\Delta q^{\rm NS}(x,Q_0^2)&=N\left(\alpha_{\rm NS},
\beta_{\rm NS},a_{\rm NS}\right)\eta_{\rm NS}x^{\alpha_{\rm NS}}
(1-x)^{\beta_{\rm NS}}(1+a_{\rm NS}x)\cr
\Delta g(x,Q_0^2)&=N\left(\alpha_g,
\beta_g,a_g\right)\eta_g x^{\alpha_g} (1-x)^{\beta_g}(1+a_gx),\cr}}
where $N(\alpha,\beta,a)$ is fixed by the normalization condition
\eqn\norm
{N(\alpha,\beta,a)\int_0^1\!dx\, x^{\alpha}
(1-x)^{\beta}(1+ax)=1.}
Similar parametrizations have been used in recent global fits of polarized
parton distributions~\stirge. The large $x$ behaviour of the parametrization
eq.~\parm\ is inspired by QCD counting rules~\ref\brod{See S.~J.~Brodsky,
M.~Burkardt and I.~Schmidt, preprint SLAC-PUB-6087 (1994) and ref. therein},
{}from which we expect $\beta_q=\beta_{\rm NS}\simeq3$ and $\beta_g\simeq 4$.
The polynomial in $x$ (which we take to be first order) is a phenomenological
interpolation which has the purpose of improving the quality of the fit in the
intermediate $x$ region.

Several estimates for the small $x$ behaviour of $g_1$ are available.
Arguments based on the dominance of known Regge
poles~\ref\heim{R.~L.~Heimann, \NP\vyp{B64}{1973}{429}.} lead to the
expectation
\eqn\smallxre
{ g_1\neath \sim {x\to 0} x^\lambda;\quad 0\le\lambda\le 0.5.}
However, coherence arguments~\brod\ suggest that (at a typical nucleon scale)
the polarized gluon distribution should be related to the unpolarized
one $g(x)$ according to
\eqn\smallxbro
{{\Delta g\over g}\neath \sim {x\to 0} x.}
Thus, if the behaviour of is $g$  dominated by a soft
pomeron~\ref\smallx{See \eg\ B.~Bade\l ek et al., \RMP\vyp{64}{1992}{927}.},
then  $g(x)\sim{1\over x}$ so that  the lower bound for $\lambda$ in
eq.~\smallxre\ is saturated, but if $g(x)$ has a harder behaviour
then $\lambda<0$ ($\lambda\sim -0.1$ with a supercritical pomeron,
and even lower with the Lipatov hard pomeron~\smallx, however
values as low as
$-0.5$ are now clearly ruled out by data from HERA~\das).
A model of the pomeron based on nonperturbative gluon
exchange~\ref\bass{S.~D.~Bass and P.~V.~Landshoff,
\PL\vyp{B336}{1994}{537}.}\ gives the still singular but softer behaviour
\eqn\smallxbala
{g_1\neath \sim {x\to 0}-2 \ln x.}
Finally, it has been argued~\ref\CR{F.~E.~Close and R.~G.~Roberts,
\PL\vyp{B336}{1994}{257}.} that it is possible for negative signature
cuts to induce an extremely singular behaviour
\eqn\smallxcloro
{g_1\neath \sim {x\to 0}{1\over x \ln^2 x}.}

Nonperturbative arguments do not generally distinguish between the leading
behaviour of the quark and that of the gluon.
We will hence consider several
representative cases of soft or hard boundary conditions of the form
eq.~\parm\ for both the
quark and gluon distributions.\foot{We
will not consider logarithmic corrections to the power-like behaviour of the
form~\smallxbala\ or \smallxcloro\ since the coefficient functions
already display a logarithmic rise at small $x$, and thus it would be
very hard to disentangle from the data
such corrections to the boundary conditions.}
Notice, however, that
since QCD evolution mixes the quark and gluon distributions
(as in eq.~\evecs), if either of the two
has a very singular behaviour, on propagation the stronger singularity
will dominate both. Also,
behaviours of the form of eq.~\smallxbala\ or
eq.~\smallxre\ with $\lambda\lsim 0.3$
will grow rapidly upon evolution according to eq.~\asuniv\
and thus their precise form will be
hidden by the perturbatively generated rise of $g_1$.

We choose the starting scale at which the parametrization eq.~\parm\
is imposed to be $Q^2_0=1$~GeV$^2$.
Here, this has the advantage that $Q^2_0<Q^2_{\rm exp}(x)$ for
all $x$, so that we may compare to the data without having to
evolve $g_1$ backwards in $Q^2$ (which
would in general be unstable, due to the causal nature of the
Altarelli-Parisi equations). Of course the price to be paid for this
choice is that at such a low scale there could be sizable higher
twist effects. However, in the unpolarized
case these effects seem mostly to be
concentrated in the large $x$ region~\ref\virmil{M.~Virchaux and A.~Milsztajn,
\PL\vyp{B274}{1992}{221}.}, where $Q^2_{\rm exp}(x)$ is
large. Because the evolution of the leading twist is
decoupled from that of the higher twist contributions,
the overall effect on our
results of having neglected
higher twist corrections  should be small. However, our final results
for $g_1(x,Q^2)$
should only be trusted in the region where the higher twist
effects are small, and in particular do not necessarily provide an accurate
description of the data at large $x$ ($x\gsim 0.3$) and small scales ($Q^2\sim
Q_0^2$ ).

At $Q^2=Q_0^2$, the quark distributions should be dominated by the
valence component, hence we will take $\beta_q=\beta_{\rm NS}$
and $a_q=a_{\rm NS}$. We also take for
simplicity the gluon parameters to be the same as the singlet ones, i.e.
$\beta_g=\beta_q$ and $a_g=a_q$, on the grounds that QCD
evolution will anyway level small differences in the initial values of the
various coefficients. More detailed fits could be obtained by relaxing these
assumptions or increasing the order of the polynomial in $x$; however, our
aim is not to propose a new parton parametrization, but rather to study the
scale dependence of $g_1$, especially in the small $x$ region,
where these details should make
very little difference. The accuracy of present day data is in any
case not good enough to justify fitting all these parameters
independently~\stirge.

Given that no clear-cut theoretical prediction
for the small $x$ behaviour of $\Delta q$ and $\Delta g$ at
a fixed scale exists,
while we do not expect the present data to allow fitting the
 exponents $\alpha_q$ and
$\alpha_g$, we will let them
vary independently over the range of values $-1<\alpha_q,\alpha_g\leq 0$,
and study the dependence of the results on this choice.
The range of positive values eq.~\smallxre\ all give essentially the same
behaviour of $g_1$ (because of the logarithmic
rise in the coefficient function)
and are thus all accounted for by the flat case $\alpha_q=0$, $\alpha_g=0$.
The lower bound is necessary since
the  first moment of the parton distributions must be finite.
The nonsinglet quark distribution, which does not mix with the gluon and
with the sea, will be assumed to be valence-like, i.e. to have
$\alpha_{\rm NS}\ge0$; specifically we will take $\alpha_{\rm
NS}=0.2$, which is the best-fit value of ref.~\ANR\ (where only the
earlier data in the valence region  were considered).

We are thus finally left with seven free parameters, namely
the two parameters $\beta$ and $a$ controlling the large $x$
behaviour, the two small-$x$ exponents $\alpha_q$ and $\alpha_g$,
and the three normalization coefficients
$\eta_q$, $\eta_{\rm NS}$ and $\eta_g$. The
normalization of the
nonsinglet, however, is fixed by the fact that the first moment
of $\Delta q^{\rm NS}$ is $Q^2$-independent to all orders because of
the conservation of the nonsinglet axial current. It can thus be
determined from the knowledge of the nonsinglet component of the matrix
element of the axial current  measured in hyperon $\beta$  decays:
\eqn\normns
{\eta_{\rm NS}=
\int_0^1 \Delta q^{\rm NS}(x,Q^2)dx=\smallfrac{3}{4}g_A+\smallfrac{1}{4}a_8,}
where~\CR
\eqn\nsval
{\eqalign{g_A&=1.2573\pm 0.0028\cr
a_8&=0.579\pm 0.025.\cr}}

The normalization of the gluon and singlet quark contributions to $g_1$
cannot be fixed analogously, due to the anomaly which afflicts the singlet
axial curent.
At a low scale, such as $Q^2\sim Q_0^2$, $\eta_q$ might
be expected to be close to
the normalization of the polarized quark distribution (i.e., the
total fraction of spin carried by quarks) as calculated in quark
models. One may estimate this from the normalization of the octet
current $a_8$ eq.~\normns, which for valence
quarks according to the Zweig rule will coincide with that of the
singlet. The difference between $\eta_q$ and the observed small
value of $\Gamma_1$ (which is much larger than the evolution effects)
is then be made up by a large value of $\eta_g$~\alros. This
interpretation of the data, however appealing, need not
necessarily be true; we  will therefore consider three different scenarios:
i) a ``maximal'' gluon case, namely $\eta_q=a_8$; ii) the opposite limit
where $\eta_g=0$ and any polarized gluons are generated dynamically;
iii) a more general fit where both $\eta_q$ and $\eta_g$ are left as free
parameters. Note that rather than using the experimental
determinations of the first moment of $g_1$ as a constraint on
$\eta_q$ and $\eta_g$, as in ref.~\stirge, we will instead determine
$\Gamma_1$ using our own fits to the $g_1$ data, evolved to a common scale.

The results of the fit to the data are displayed in table 1 and figure 2 (in
several typical cases) and can now be used to draw some quantitative
conclusions on the scale dependence and small-$x$ behaviour of $g_1$. First, it
is clear that a statistically significant evolution effect is seen in the data.
Indeed, throughout the intermediate $x$ region $0.03\le x\le 0.2$, where both
experiments have data, there would appear to be a systematic discrepancy in
that all of the E143 data~\Eoft\  lie  systematically about one standard
deviation  below the SMC data~\SMC. The difference is however seen to be
entirely accounted for by the fact  that in each $x$ bin the E143 data are
taken at a smaller value of $Q^2$ than the SMC data. This is displayed
explicitly in fig.~2a-c: in particular, fig.~2c shows that even if $\eta_g=0$
and $\alpha_q=0$, i.e. if the small $x$ behaviour of the initial conditions  is
very soft, the dynamically generated gluon polarization is sufficient to drive
the required amount of perturbative evolution. Notice that (as shown explicitly
by the calculation of ref.~\ANR) not enough evolution would be found if we had
set $C_g=0$ in eq.~\gone. In fact, the total $\chi^2$ of the fit shown in
fig.~2c (which is the least sensitive to this effect since it has minimal
gluon) increases by more than three units if we set the coefficient functions
to their one loop values eq.~\llogc. This provides a posteriori evidence in
favor of our approach, i.e. it supports the assumption that the inclusion of
the next-to-leading order coupling of $\Delta g$ to $g_1$ even when the
Altarelli-Parisi equations are only solved at leading order improves the
determination of $g_1$; this can be thus interpreted, albeit with some caution,
as direct experimental evidence for the anomaly-induced contribution of $\Delta
g$ to $g_1$.

Even though the evolution of $\Gamma_1$ is fixed in a universal way by the
anomalous dimension of the axial current $j^\mu_5$ (and is slight, since it
starts at two loops), the evolution of $g_1$ for any finite value or range of
$x$ is a leading order (potentially large) effect. However, it is not
universal, but rather depends on the specific form of $\Delta q$ and $\Delta
g$. In practice, the parameter which controls the exact amount of evolution is
essentially the value of $\eta_g$, i.e., the size the gluon contribution at the
starting scale. Large values of $\eta_g$ lead to significantly larger
evolution. The value of $\eta_g$ cannot be determined from the data, but may be
somewhat constrained. The limit on $\eta_g$ may be read off table~1; its actual
value depends on the values of $\alpha_q$ and $\alpha_g$. In general, the data
favor intermediate values of $\eta_g$, i.e. such that $\eta_g>0$, but still not
large enough that the condition $\eta_q=a_8$ be satisfied at the starting
scale.

On the contrary, no significant constraint is put by the data on the values of
the small $x$ exponents $\alpha_q$, $\alpha_g$, as expected from fig.~1. Thus a
good fit can be obtained  with either ``hard'' and ``soft'' boundary
conditions, such as, for example $\alpha_q=\alpha_g=-0.5$ and
$\alpha_q=\alpha_g=0$ (see fig.~2a,~b). In fact, with all combinations of
small-$x$ exponents it is possible to accurately fit the data (table 1a) by
suitably adjusting the value of $\eta_g$.  However, for different classes of
values of $\alpha_q$ and $\alpha_g$ the qualitative form of the perturbative
evolution in the intermediate $x$  range ($0.03\le x\le0.1$) may be
significantly different. In particular, if $\alpha_g<\alpha_q$ (gluon more
singular than quark) $g_1$ becomes rapidly large and negative at small $x$. It
follows that only small values of $\eta_g$ (fig.~2a) are allowed so that the
decrease of $g_1$ falls outside the observed range. If instead
$\alpha_q\le\alpha_g$ (quark at least as singular as gluon), then the onset of
the small-$x$ asymptotic behaviour (where $\Delta G$ and $\Delta q$ are
anticorrelated, according to eq.~\evecs) tends to generate oscillatory
interference patterns in $g_1$ (fig.~2d). These oscillations are more
pronounced if $\eta_g$ is large (hence the evolution is stronger), and for soft
boundary conditions (i.e. $\alpha_q\approx\alpha_g\gsim - 0.3$), when the
asymptotic double logarithmic growth eq.~\asuniv\ only sets in slowly. This
also allows us to rule out a maximal gluon for soft boundary conditions, since
then fluctuations would be too large (fig.~2e). Finally, if $\alpha_q$ is very
large, then it is possible to satisfy the condition $\eta_q=a_8$ even though in
fact $\eta_g$ is very small (compatible with zero): in such case $g_1$ grows
very large in the unobserved small-$x$ region. This
scenario~\ref\clsin{F.~E.~Close and R.~G.~Roberts, \PRL\vyp{60}{1988}{1471}.}
is also compatible with the data (fig.~2f): an arbitrarily large contribution
to $\Gamma_1$ concentrated at very small $x$ does not influence the evolution
at larger $x$ and thus cannot be seen in  $g_1(x)$ measured over any finite
region of $x$.

Conversely, if $\eta_g$ is fixed (for example on the basis of theoretical
prejudice), then the values of $\alpha_q$, $\alpha_g$ can be fitted (or at
least constrained). In particular, if $\eta_g$ is very large (maximal), then
only values $\alpha_q,\alpha_g<-0.5$, with $\alpha_q<\alpha_g$ are allowed, the
corresponding minimum of the $\chi^2$ in the $(\alpha_q,\alpha_g)$ plane being
rather narrow. If $\eta_g=0$, then a valence-like quark is favored
($\alpha_q\approx +0.5$) , albeit with a sizable uncertainty. For intermediate
values of $\eta_g$ the $\chi^2$ is very smooth in the $(\alpha_q,\alpha_g)$
plane, and $\alpha_q$ and $\alpha_g$ are essentially unconstrained.

In order to check that these results are not a byproduct of our treatment of
next-to-leading corrections (which are included in the coefficient functions
but not in the Altarelli-Parisi splitting functions), we have repeated our fits
with the two alternative forms of gluon coefficient functions given in
ref.~\stirge. The results  do not change in a statistically significant way.

\newsec{The First Moment}

On the basis of the quantitative knowledge on the evolution of $g_1$ acquired
in the previous section, we may now discuss the values of its first moment
$\Gamma_1$ eq.~\fmom, and in particular the contribution to it from the small
$x$ region
\eqn\fmsx
{\Gamma_1^{sx}(Q^2)=\int_0^{0.01} \!dx\, g_1(x,Q^2).}
Values of $\Gamma_1$ and $\Gamma_1^{sx}$ at various scales, and the value of
the singlet contribution $a_0(Q^2)$ eq.~\axfmom\  are given in table 2, for
several of  the choices of the fit parameters which give good agreement with
the data.

The contribution eq.~\fmsx\ of the small $x$ extrapolation to $\Gamma_1$, at
the scale of the data, turns out to be approximately the same for all
reasonable fits (apart of course for the extremely singular  case of fig.~2f).
The uncertainty in the knowledge of the small-$x$ behaviour of the starting
quark and gluon distributions eq.~\parm\ is thus seen to have very little
effect on the value of $\Gamma_1$.

However, there is a hitherto unnoticed sizable uncertainty in the value of
$\Gamma_1$ at any given scale which is only indirectly related to the small $x$
extrapolation and is due to the scale dependence of $g_1$. This may be seen by
comparing, for instance, the results of fitting the data with $\eta_g=0$ and
$\alpha_q=0$ (fig.~2c), or with $\alpha_q=-0.5$, $\alpha_g=0$ and
maximal gluon, $\eta_q=a_8$ (fig.~2d). These fits have comparable $\chi^2$,
thus providing an equally accurate description of the data, and the same value
of $\Gamma_1^{sx}(Q^2)$ at $Q^2=3$~GeV$^2$. Nevertheless they lead to
values of $\Gamma_1(Q^2)$ which differ by almost 10\%, hence to values of its
singlet component which differ by almost a factor 2.

The origin of this uncertainty can be understood by remembering that whereas
the singlet contribution $a_0$ eq.~\axfmom\ to $\Gamma_1$ is scale independent
at leading order, the integral of $g_1$ restricted to a limited $x$ region is
not. Inspection of the results of table~2 and fig.~2 shows that, for
$\eta_g>0$, in the intermediate $x$ range ($0.03\lsim x \lsim 0.2$) $g_1$, and
thus the contribution to $\Gamma_1$ from this region, is a rather rapidly
increasing function of $Q^2$; this is also manifestly displayed by the data.
This implies that the contribution $\Gamma_1^{sx}$ from the (unmeasured) small
$x$ region must decrease as $Q^2$ increases. However, this $Q^2$ dependence of
$\Gamma_1^{sx}$ is extremely sensitive to the value of $\eta_g$: if $\eta_g$ is
large, then $\Gamma_1^{sx}$ decreases very rapidly, due to the large
oscillations induced in $g_1$, while this is not the case if $\eta_g$ is small.
But the value of $\eta_g$ is only weakly constrained by the data, since the
corresponding evolution effects are mostly concentrated in the experimentally
unaccessible small $x$ region. This implies that if $\eta_g$ is large, then
$\Gamma_1$ can be substantially smaller at the starting scale  because $g_1$ in
the intermediate $x$ region, which provides the bulk of it, has been evolved
more.

As a corollary to this result, we find that our determinations of $\Gamma_1$
are systematically smaller than those quoted in refs.~\refs{\SMC,\Eoft}. This
is because the data points which provide the bulk of $\Gamma_1$ are taken at a
scale $Q^2$ larger than the nominal average scale of either experiment: for
example in ref.~\SMC\ $Q^2_{\rm exp}(x)$ is always larger than $20$~GeV$^2$ for
$x$ above $0.1$,  while $\langle Q^2\rangle=10$~GeV$^2$ for this experiment.
Since in this region $g_1$ increases as the scale increases the approximate
treatment of the scale dependence in refs.~\refs{\SMC,\Eoft}\ leads to an
overestimate of $\Gamma_1$.\foot{Notice however that part of the effect is due
to the fact that our determination of $g_1$ from the data of ref.~\Eoft\ is
systematically smaller than that presented there, because of the
kinematic higher twist effects included in their determination of $g_1$
 from the measured asymmetry.} Conversely, the small-$x$ extrapolation of
ref.~\SMC\ overestimates the data, because it is obtained by extrapolating the
smallest $x$ data points, which are taken at very low $Q^2$ ($\sim 1$~GeV$^2$),
and then assumed to apply at $<Q^2>=10$~GeV$^2$; this turns out not to be the
case for ref.~\Eoft\ partly because their value of $<Q^2>=3$~GeV$^2$
is closer to that of the small $x$ bins, and partly just because
a straight line drawn from the smallest $x$ points of this
experiment happens to be closer to our results.

This analysis is summarized pictorially in fig.~3, where we display the scale
dependence of both $\Gamma_1$ and the small $x$ contribution to it,
$\Gamma_1^{sx}$, for representative minimal, intermediate, and maximal values
of $\eta_g$. Clearly, $\Gamma_1^{sx}$ decreases significantly as a function of
scale if $\eta_g$ is maximal, whereas it is essentially scale-independent if it
is minimal (the slight rise displayed in the figure in such case is not
significant to the level of accuracy of the present treatment: it could be
turned into a decrease for different forms of the next-to-leading corrections
which we do not include). As a consequence, larger values of $\eta_g$
correspond to smaller values of $\Gamma_1$. Since however the value of $\eta_g$
cannot be fixed on the basis of theoretical arguments, and is not constrained
by the data, this is to be considered as an intrinsic uncertainty on the
determination of $\Gamma_1$. However, even with minimal values of $\eta_g$,
inclusion of the next-to-leading evolution effects discussed here leads to a
smaller value of $\Gamma_1$ than that found through the approximate treatment
of the scale dependence given by the experimental collaborations
\refs{\SMC,\Eoft}.

\newsec{Discussion}

The main result of our analysis is that the next-to-leading order coupling of
the polarized gluon distribution to $g_1$ has a sizable effect on the scale
dependence of the structure function. This effect has important consequences
for the extraction of the first moment $\Gamma_1$ of $g_1$ from the data
because, even though $\Gamma_1$ is scale-independent at leading order, the
contribution to it from any finite $x$ range (such as the experimentally
accessible range) is not. In particular, the contribution to $\Gamma_1$ from
the unobserved small-$x$ region turns out to be a decreasing function of the
scale, this decrease being compensated by an approximately equal increase of
the contribution from the measured region. The importance of the evolution
effects is largely independent of the detailed shape of the quark and gluon
distributions, but increases with the size of the polarized gluon distribution.

As a consequence, due to the neglect of these evolution effects, the value of
$\Gamma_1$ is somewhat overestimated by present experiments, and the
uncertainty, which is due to the lack of knowledge of the size of the polarized
gluon, is substantially underestimated. 
more accurate approximate 
$<Q^2>$ closer to the value of $Q_{\rm exp}^2(x)$ in the $x$ region 
gives the dominant contribution to $\Gamma_1$. Indeed, this 
considerably in reducing the uncertainty on the 
region. The largest uncertainty would then 
contribution which would be evolved 
uncertainty in the knowledge of the small $x$ behaviour of polarized parton
distributions leads to a comparatively much smaller ambiguity, at least
excluding the possibility of a very large contribution to $g_1$ from the very
small $x$ region (such as a delta-like contribution). A more accurate
determination of $\Gamma_1$ and especially of the associated error can only be
obtained by including the full scale dependence in the analysis of the data.
 From table~2 we suggest the value
\eqn\salame
{\Gamma_1(Q^2=3\;{\rm GeV}^2)=0.120\pm0.005\pm0.005}
where the first error is statistical and the second is an optimistic estimate
of the theoretical uncertainty in the scale dependence due to the uncertainty
in the size of the gluon distribution. The systematic error, which is typically
of order $0.010$, must still be added.

The intrinsic ambiguity related to evolution could be substantially reduced if
more precise data in the intermediate $x$ region ($0.03\lsim x\lsim 0.2$) were
available for somewhat different values of $Q^2$: because of the  instability
of $g_1$ under perturbative evolution even a moderate improvement in
experimental knowledge could result in much more stringent theoretical
constraints on the size of the polarized gluon distribution, which in turn
would be enough to determine the small $x$ behaviour of $g_1$ to much better
accuracy. Such information would thus be more conclusive than direct
information on $g_1$ in the small $x$ region itself. The determination even of
the qualitative behaviour of $g_1$ at small $x$, in the same way as in  the
unpolarized case~\das\ would instead only be feasible  with a kinematic
coverage of the small $x$ region comparable to that which HERA data provide at
present for $F_2$.

In short, the scale dependence of $g_1$ induced by the axial anomaly turns out
to be an essential ingredient in its phenomenology: a more detailed
understanding of it will only be possible once the actual size of the gluon
polarization is known more accurately. On the theoretical side, the
determination of the full set of two loop Altarelli-Parisi splitting function
will allow a fully consistent treatment. It will then perhaps be possible to
extract information on the polarized gluon distribution, which is of
considerable intrinsic theoretical interest, directly from structure function
data even over a moderate range of $x$ and $Q^2$.

\bigskip
{\bf Acknowledgement:} We thank G.~Altarelli for discussions, and P.~Nason and
S.~Rollet for useful suggestions.
\vfill
\eject

\midinsert\hfil
\vbox{\tabskip=0pt \offinterlineskip
      \def\tablerule{\noalign{\hrule}}
      \halign to 412pt{\strut#&\vrule#\tabskip=1em 
                   &\hfil#&\vrule#
                   &\hfil#&\vrule#
                   &\hfil#&\vrule#
                   &\hfil#&\vrule#
                   &\hfil#\hfil&\vrule#
                   &\hfil#\hfil&\vrule#
                   &#\hfil&\vrule#\tabskip=0pt\cr\tablerule
             &&\omit\hidewidth $\alpha_q$\hidewidth
             &&\omit\hidewidth $\alpha_g$\hidewidth
             &&\omit\hidewidth $\beta$\hidewidth
             &&\omit\hidewidth $a$\hidewidth
             &&\omit\hidewidth $\eta_q$\hidewidth
             &&\omit\hidewidth $\eta_g$\hidewidth
             &&\omit\hidewidth $\chi^2$\hidewidth&\cr\tablerule
             &&\multispan{13}  Fitted gluon:\hfil&\cr\tablerule
&&$ 0.0$&&$ 0.0$&&$2.95\pm 0.50$&&$1.44\pm 1.93$
                &&$0.18\pm 0.88$&&$0.22\pm 0.50$&&$ 31.0^{\rm ~}$&\cr
&&$ 0.0$&&$-0.2$&&$2.94\pm 0.47$&&$1.43\pm 1.74$
                &&$0.18\pm 0.72$&&$0.22\pm 0.42$&&$ 30.9^{\rm b)}$&\cr
&&$ 0.0$&&$-0.5$&&$2.98\pm 0.49$&&$1.62\pm 1.96$
                &&$0.16\pm 0.73$&&$0.18\pm 0.51$&&$ 30.9^{\rm ~}$&\cr
&&$ 0.0$&&$-0.9$&&$3.07\pm 0.38$&&$2.11\pm 1.59$
                &&$0.15\pm 0.45$&&$0.09\pm 1.58$&&$ 31.1^{\rm ~}$&\cr
&&$-0.2$&&$ 0.0$&&$2.96\pm 0.49$&&$1.57\pm 1.97$
                &&$0.21\pm 0.11$&&$0.35\pm 0.59$&&$ 31.0^{\rm ~}$&\cr
&&$-0.5$&&$ 0.0$&&$2.99\pm 0.48$&&$1.78\pm 2.12$
                &&$0.32\pm 0.19$&&$0.74\pm 0.86$&&$ 31.0^{\rm ~}$&\cr
&&$-0.9$&&$ 0.0$&&$3.45\pm 0.26$&&$5.83\pm 2.36$
                &&$0.43\pm 1.00$&&$0.00\pm 1.12$&&$ 33.0^{\rm ~}$&\cr
&&$-0.2$&&$-0.2$&&$2.94\pm 0.46$&&$1.54\pm 1.76$
                &&$0.20\pm 0.87$&&$0.35\pm 0.48$&&$ 30.8^{\rm ~}$&\cr
&&$-0.5$&&$-0.5$&&$2.98\pm 0.40$&&$1.91\pm 1.64$
                &&$0.28\pm 0.10$&&$0.63\pm 0.54$&&$ 30.4^{\rm a)}$&\cr
&&$-0.9$&&$-0.9$&&$3.12\pm 0.33$&&$3.00\pm 1.70$
        &&$0.97\pm 0.33$&&$2.80\pm 1.58$&&$ 29.8^{\rm ~}$&\cr\tablerule
             &&\multispan{13}  Maximal gluon:\hfil&\cr\tablerule
&&$ 0.0$&&$ 0.0$&&$2.17\pm 0.34$&&$-0.61\pm 0.30$
                &&$0.58$&&$2.02\pm 0.17$&&$  72.0^{\rm ~}$&\cr
&&$ 0.0$&&$-0.2$&&$2.11\pm 0.34$&&$-0.62\pm 0.31$
                &&$0.58$&&$1.96\pm 0.18$&&$  83.2^{\rm e)}$&\cr
&&$ 0.0$&&$-0.5$&&$1.98\pm 0.35$&&$-0.62\pm 0.32$
                &&$0.58$&&$1.99\pm 0.20$&&$ 104.7^{\rm ~}$&\cr
&&$ 0.0$&&$-0.9$&&$1.76\pm 0.42$&&$-0.66\pm 0.37$
                &&$0.58$&&$4.48\pm 0.57$&&$ 141.5^{\rm ~}$&\cr
&&$-0.2$&&$ 0.0$&&$2.08\pm 0.42$&&$-0.61\pm 0.39$
                &&$0.58$&&$2.08\pm 0.17$&&$  51.1^{\rm ~}$&\cr
&&$-0.5$&&$ 0.0$&&$2.41\pm 0.53$&&$ 0.02\pm 0.86$
                &&$0.58$&&$1.87\pm 0.18$&&$  33.2^{\rm d)}$&\cr
&&$-0.9$&&$ 0.0$&&$3.42\pm 0.26$&&$ 5.30\pm 2.10$
                &&$0.58$&&$0.24\pm 0.17$&&$  33.3^{\rm f)}$&\cr
&&$-0.2$&&$-0.2$&&$2.02\pm 0.42$&&$-0.61\pm 0.39$
                &&$0.58$&&$2.04\pm 0.18$&&$  59.2^{\rm ~}$&\cr
&&$-0.5$&&$-0.5$&&$2.17\pm 0.61$&&$-0.20\pm 0.85$
                &&$0.58$&&$2.05\pm 0.20$&&$  40.5^{\rm ~}$&\cr
&&$-0.9$&&$-0.9$&&$3.36\pm 0.26$&&$ 4.56\pm 1.95$
        &&$0.58$&&$0.98\pm 0.51$&&$  31.2^{\rm ~}$&\cr\tablerule
             &&\multispan{13}  Minimal gluon:\hfil&\cr\tablerule
&&$ 0.0$&&---&&$3.09\pm 0.33$&&$2.23\pm 1.29$
             &&$0.14\pm 0.33$&&$0$&&$  31.1^{\rm c)}$&\cr
&&$-0.2$&&---&&$3.15\pm 0.31$&&$2.72\pm 1.39$
             &&$0.15\pm 0.34$&&$0$&&$  31.3^{\rm ~}$&\cr
&&$-0.5$&&---&&$3.26\pm 0.29$&&$3.76\pm 1.64$
             &&$0.17\pm 0.39$&&$0$&&$  31.7^{\rm ~}$&\cr
&&$-0.9$&&---&&$3.45\pm 0.26$&&$5.83\pm 2.36$
             &&$0.43\pm 1.00$&&$0$&&$  33.0^{\rm ~}$&\cr\tablerule
}}
\hfil\bigskip
\centerline{\vbox{\hsize= 380pt \raggedright\noindent\footnotefont
Table~1: Best-fit parameters and $\chi^2$ for the fit of $g_1$ to the
data of refs.~\refs{\SMC,\Eoft}: the letters a)--f) refer to the plots
in fig.~2.
}}
\bigskip
\endinsert
\vfill\eject

\midinsert\hfil
\vbox{\tabskip=0pt \offinterlineskip
      \def\tablerule{\noalign{\hrule}}
      \halign to 403pt{\strut#&\vrule#\tabskip=0.5em 
                   &\hfil#&\vrule#
                   &\hfil#&\vrule#
                   &\hfil#&\vrule#
                   &\hfil#&\vrule#
                   &\hfil#&\vrule#
                   &\hfil#&\vrule#
                   &\hfil#&\vrule#
                   &\hfil#&\vrule#\tabskip=0pt\cr\tablerule
       &&\omit&&\omit\hidewidth $\Gamma_1(1)$\hidewidth
             &&\omit\hidewidth $\Gamma_1(3)$\hidewidth
             &&\omit\hidewidth $\Gamma_1(10)$\hidewidth
             &&\omit\hidewidth $\Gamma_1^{\rm sx}(1)$\hidewidth
             &&\omit\hidewidth $\Gamma_1^{\rm sx}(3)$\hidewidth
             &&\omit\hidewidth $\Gamma_1^{\rm sx}(10)$\hidewidth
             &&\omit\hidewidth $a_0(3)$\hidewidth&\cr\tablerule
             &&\multispan{15}  Fitted gluon:\hfil&\cr\tablerule
&& a)&&$ 0.108\pm 0.006$&&$ 0.118\pm 0.006$&&$ 0.123\pm 0.006$
    &&$ 0.007$&&$ 0.003$&&$ 0.001$&&$ 0.102\pm 0.060$&\cr
&& b)&&$ 0.110\pm 0.005$&&$ 0.119\pm 0.005$&&$ 0.123\pm 0.005$
    &&$ 0.005$&&$ 0.004$&&$ 0.004$&&$ 0.111\pm 0.054$&\cr\tablerule
             &&\multispan{15}  Maximal gluon:\hfil&\cr\tablerule
&& d)&&$ 0.101\pm 0.005$&&$ 0.116\pm 0.005$&&$ 0.122\pm 0.005$
    &&$ 0.015$&&$ 0.005$&&$ -0.004$&&$ 0.074\pm 0.051$&\cr
&& f)&&$ 0.147\pm 0.005$&&$ 0.158\pm 0.005$&&$ 0.162\pm 0.005$
    &&$ 0.041$&&$ 0.042$&&$ 0.042$&&$ 0.496\pm 0.050$&\cr\tablerule
             &&\multispan{15}  Minimal gluon:\hfil&\cr\tablerule
&& c)&&$ 0.113\pm 0.003$&&$ 0.122\pm 0.003$&&$ 0.125\pm 0.003$
    &&$ 0.005$&&$ 0.006$&&$ 0.007$&&$ 0.136\pm 0.031$&\cr\tablerule
}}
\hfil\bigskip
\centerline{\vbox{\hsize= 380pt \raggedright\noindent\footnotefont
Table~2: Values of the first moment $\Gamma_1(Q^2)$ eq.~\fmom, at
$Q^2=1,~3,~10~\GeV^2$,
its small $x$ component eq.~\fmsx, and the matrix element $a_0$ of the singlet
axial current for several typical fits from table~1: the letters
a)--f) refer to the plots
in fig.~2.
}}
\bigskip
\endinsert
\vfill\eject
\listrefs
\vfill
\eject
\centerline{\bf Figure Captions}
\bigskip

\item{Fig.~1} Contour plot of $\ln g_1$ computed from the small $x$
evolution equation  eq.~\weq\
in the plane of the variables defined in
eq.~\xizeta, with $Q_0=1$~GeV and $x_0=0.1$. The polarized gluon and singlet
quark distributions are evolved up from  the form eq.~\parm\ with $\beta=3$,
$\eta_q=0.6$, $\eta_g=2.5$ (maximal gluon), $a=0$ and
a) $\alpha_q=0,\>\alpha_g=-0.2$ (soft); b) $\alpha_q=-0.5,\>\alpha_g=-0.5$
(hard). A scatter plot of the SMC ($+$) and E143 ($\times$)
data
is superposed for reference.
\medskip

\item{Fig.~2} Evolution of the structure function $g_1$ compared to the SMC
(crosses) and E143 (diamonds) data. The full set of parameters for the
various fits are given in table 1. a) fitted
$\eta_g$, hard boundary conditions; b) fitted $\eta_g$, soft boundary
conditions;  c)
dynamically generated gluon ($\eta_g=0$); d)
maximal gluon, hard boundary conditions; e) maximal gluon, soft boundary
conditions;
f) Ellis-Jaffe quark with small gluon contribution.
\medskip

\item{Fig.~3} Scale dependence of $\Gamma_1$ eq.~\fmom\ (a)
and the  the small $x$ ($x\le 0.01$) contribution
to it eq.~\fmsx\ (b). Solid line: fitted gluon (as in fig.~2a);
dot-dashed line: dynamically generated gluon (as in fig.~2c);
dashed line: maximal gluon (as in fig.~2d).

\vfill
\eject
\nopagenumbers
\vsize=27truecm
\null
\epsfxsize=8truecm
\hfil\epsfbox{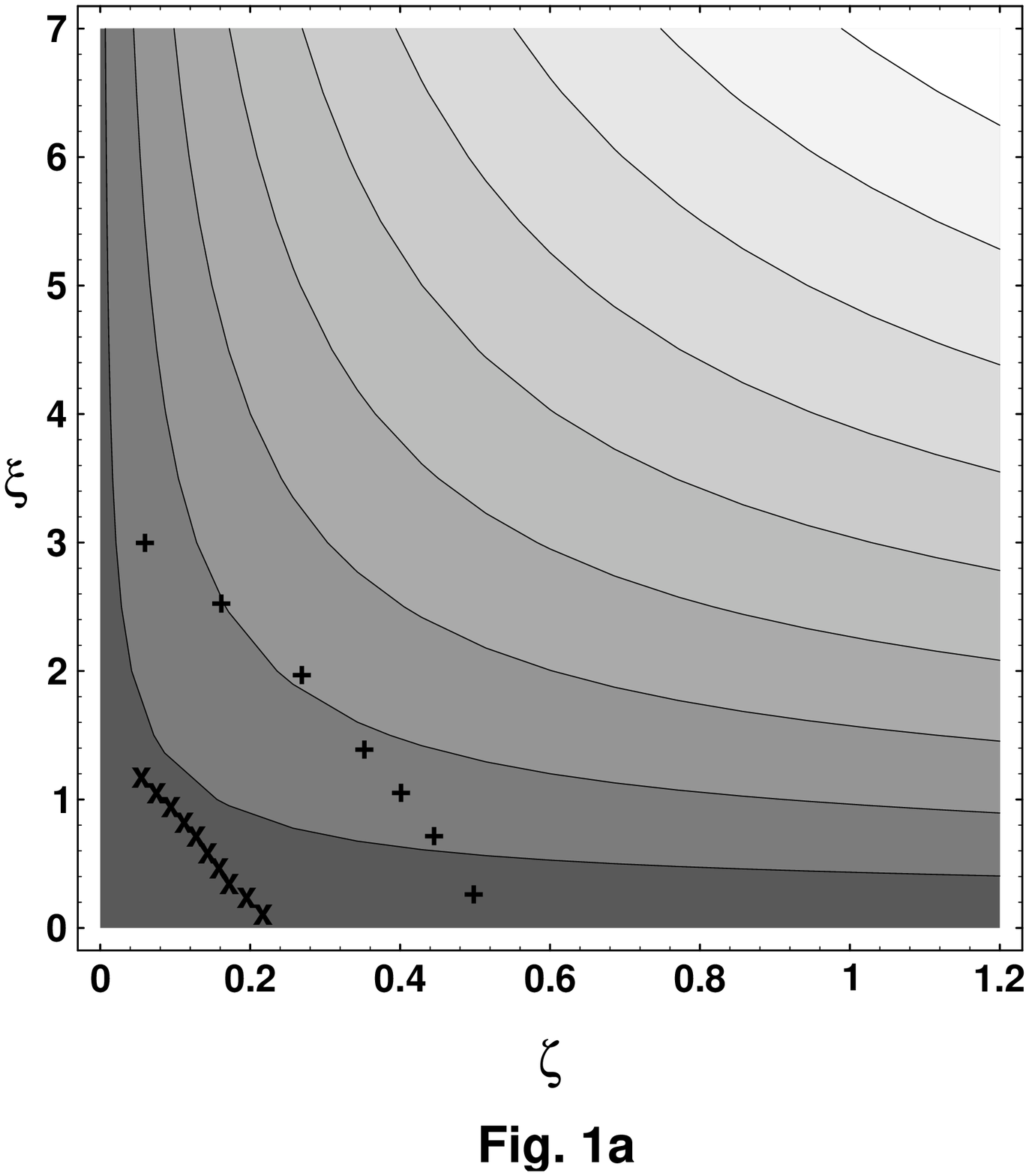}\hfil
\bigskip
\epsfxsize=8truecm
\hfil\epsfbox{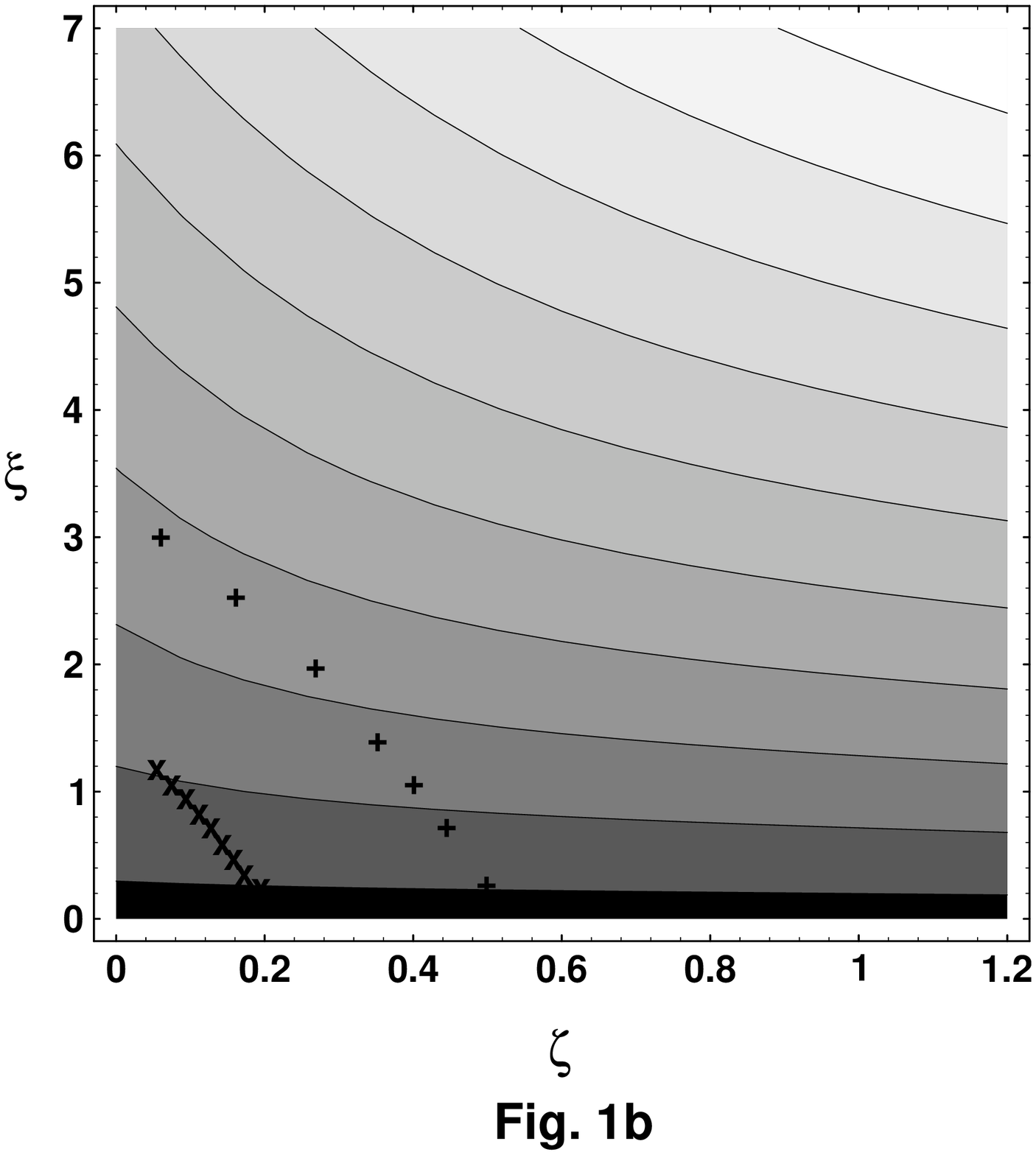}\hfil
\vfill
\eject
\null
\vskip -4.truecm
\epsfxsize=14truecm
\hfil\epsfbox{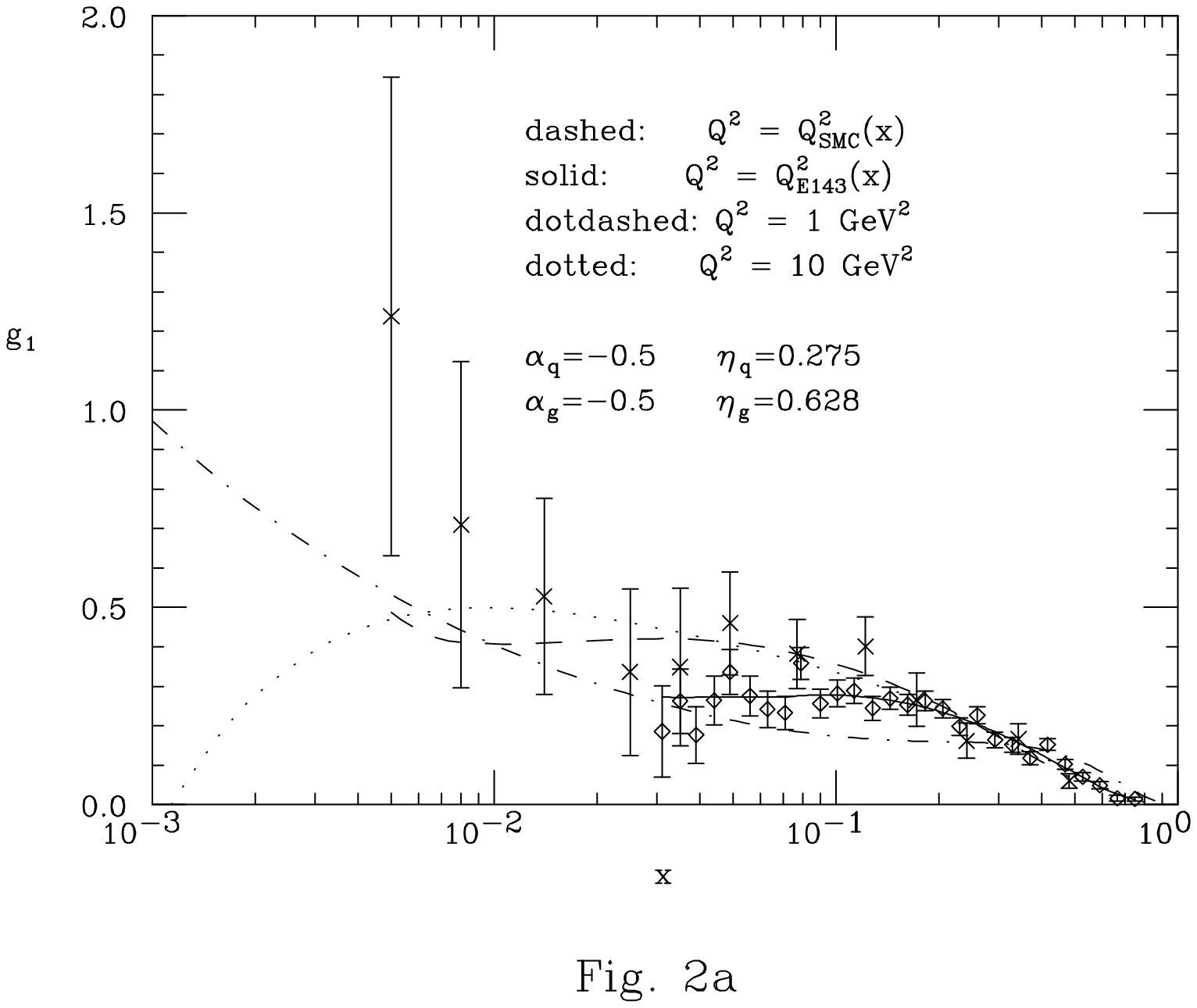}\hfil
\smallskip
\vskip -7.truecm
\epsfxsize=14truecm
\hfil\epsfbox{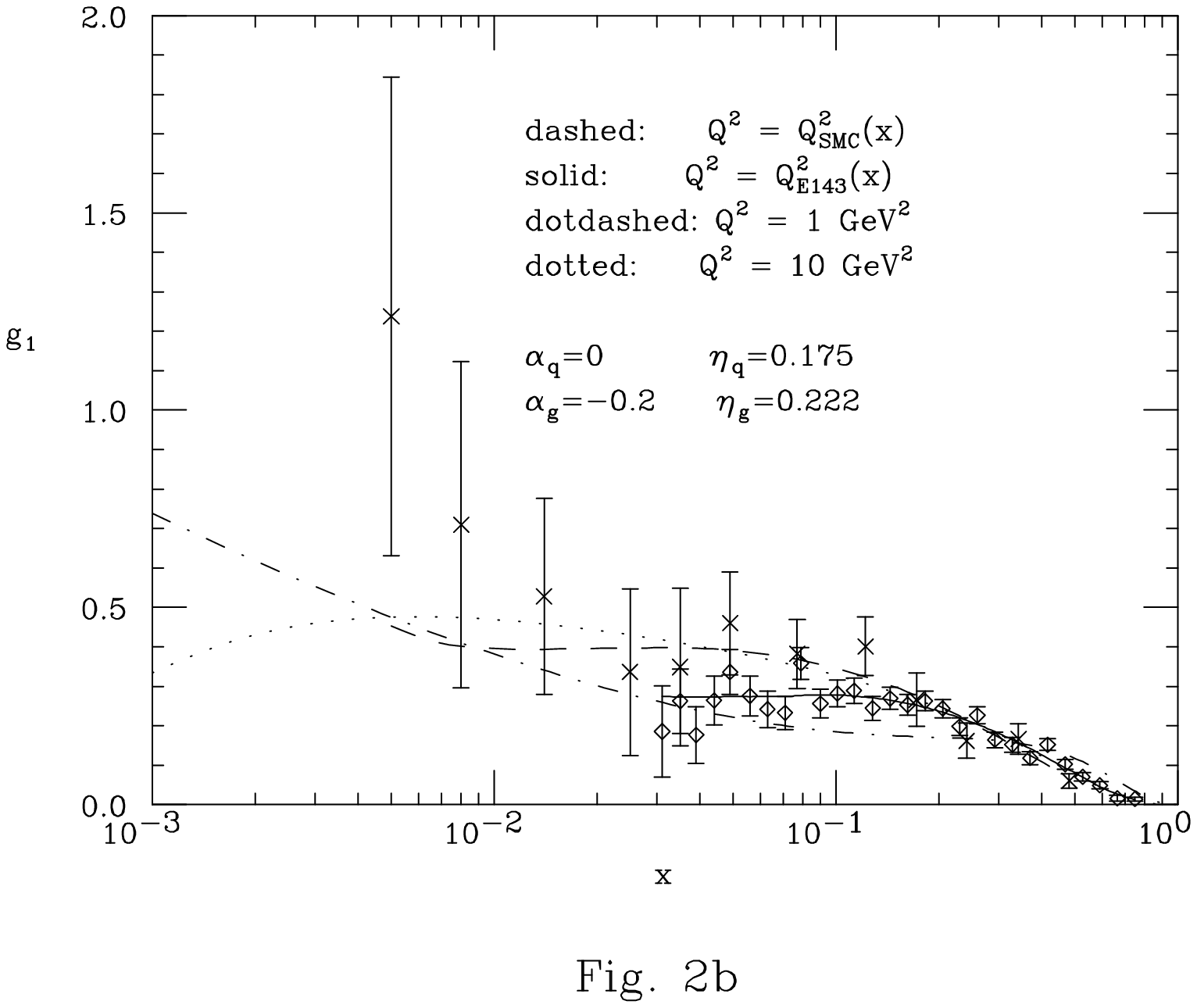}\hfil
\vfill
\eject
\null
\vskip -4.truecm
\epsfxsize=14truecm
\hfil\epsfbox{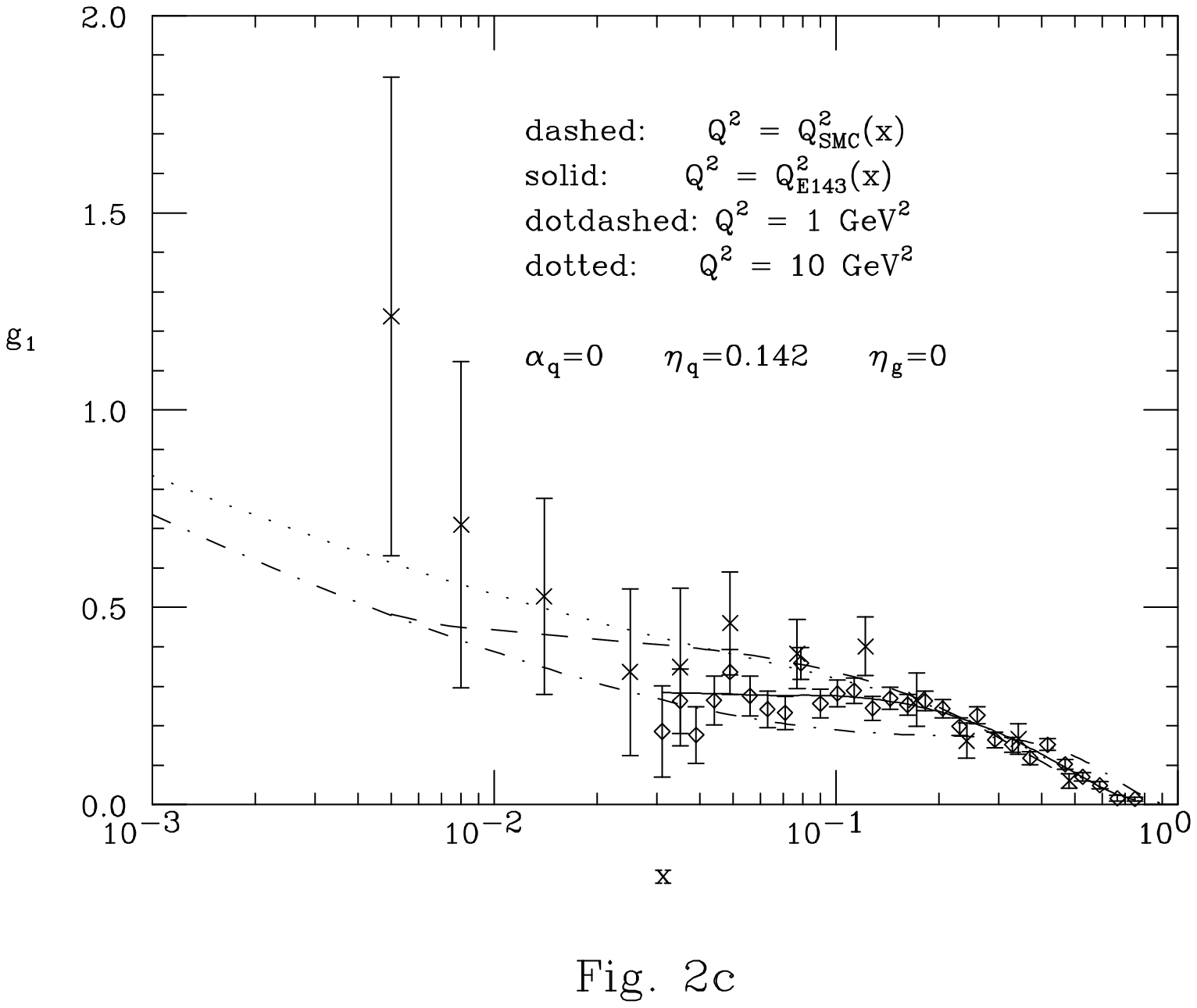}\hfil
\smallskip
\vskip -7.truecm
\epsfxsize=14truecm
\hfil\epsfbox{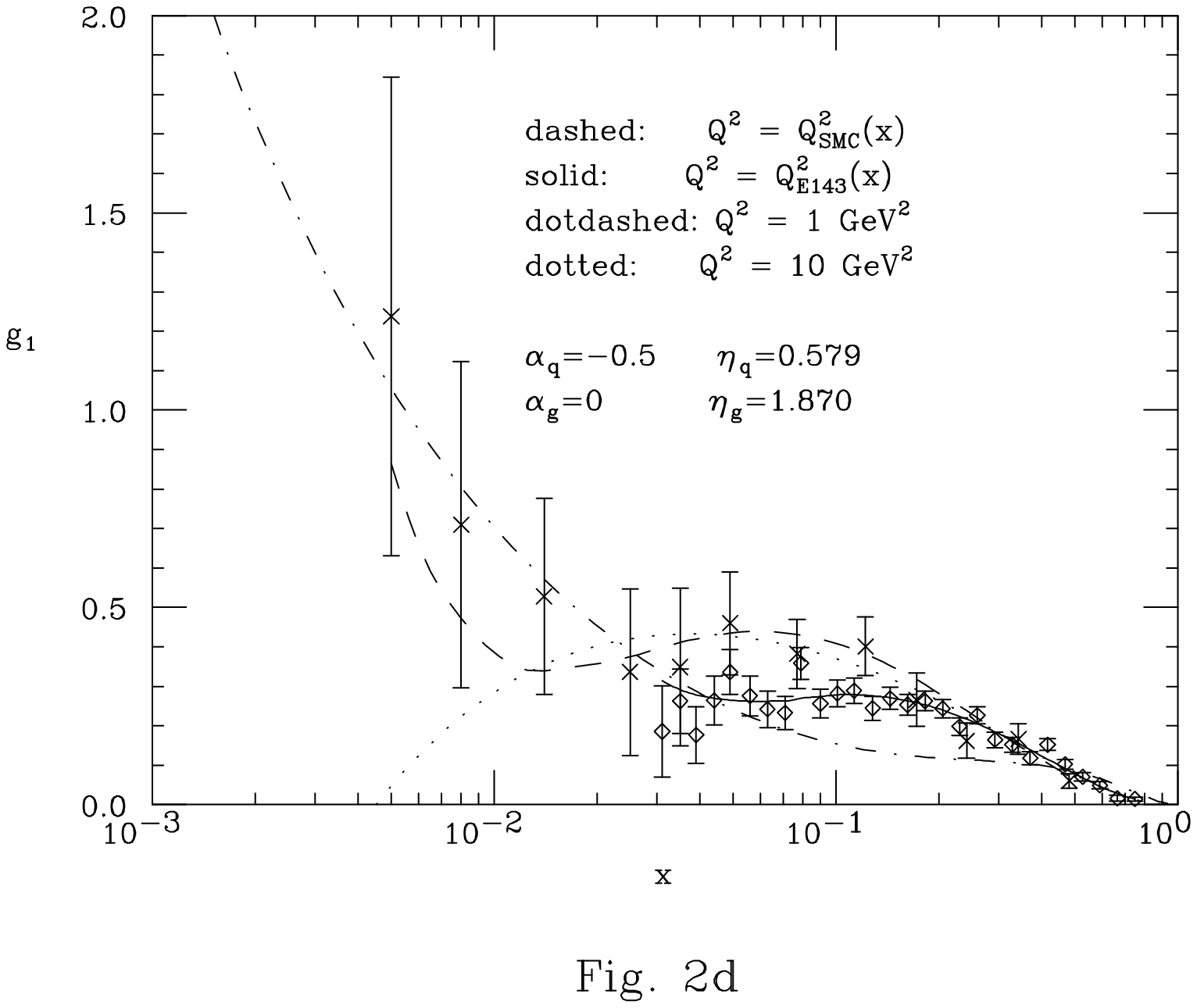}\hfil
\vfill
\eject
\null
\vskip -4.truecm
\epsfxsize=14truecm
\hfil\epsfbox{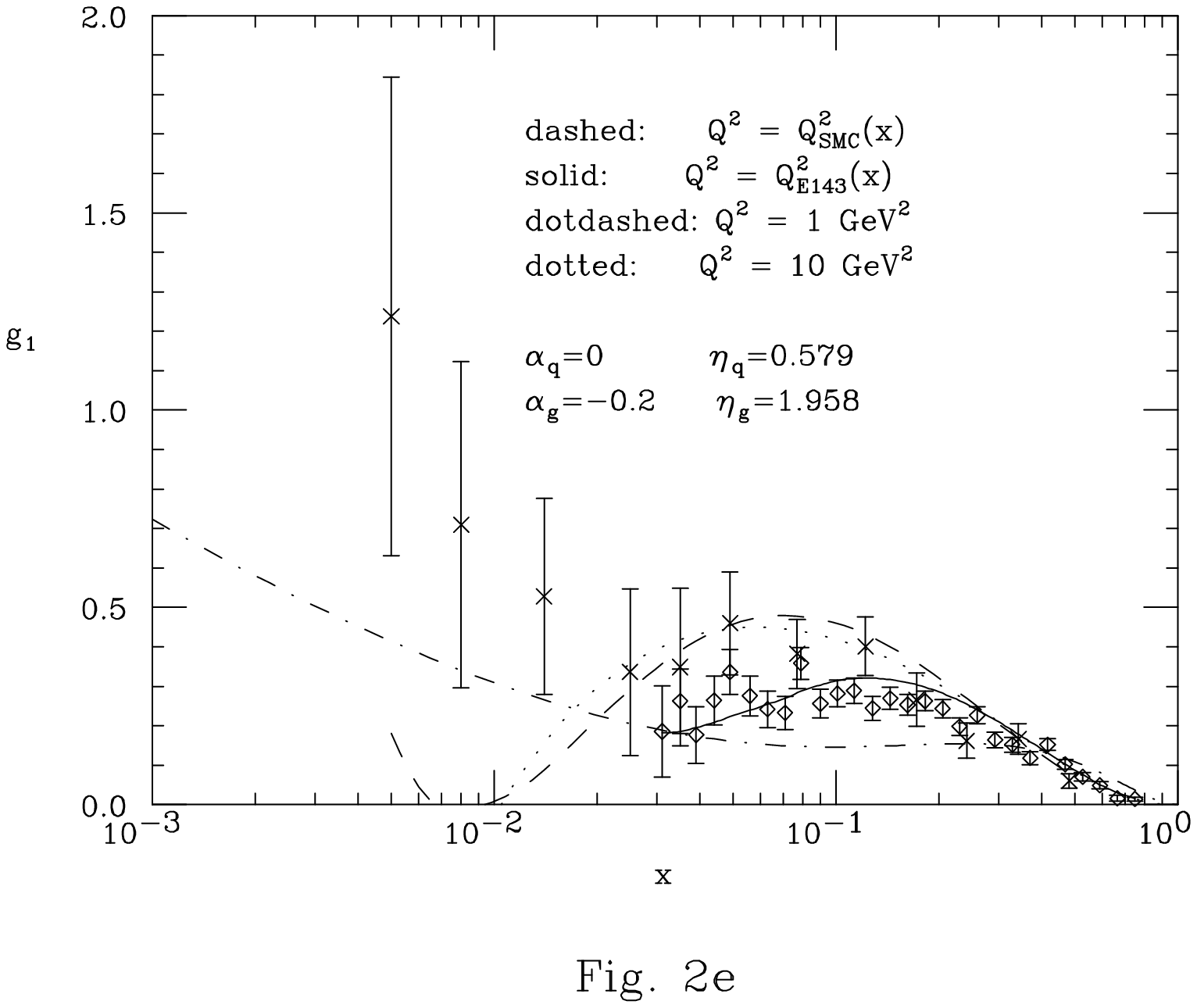}\hfil
\smallskip
\vskip -7.truecm
\epsfxsize=14truecm
\hfil\epsfbox{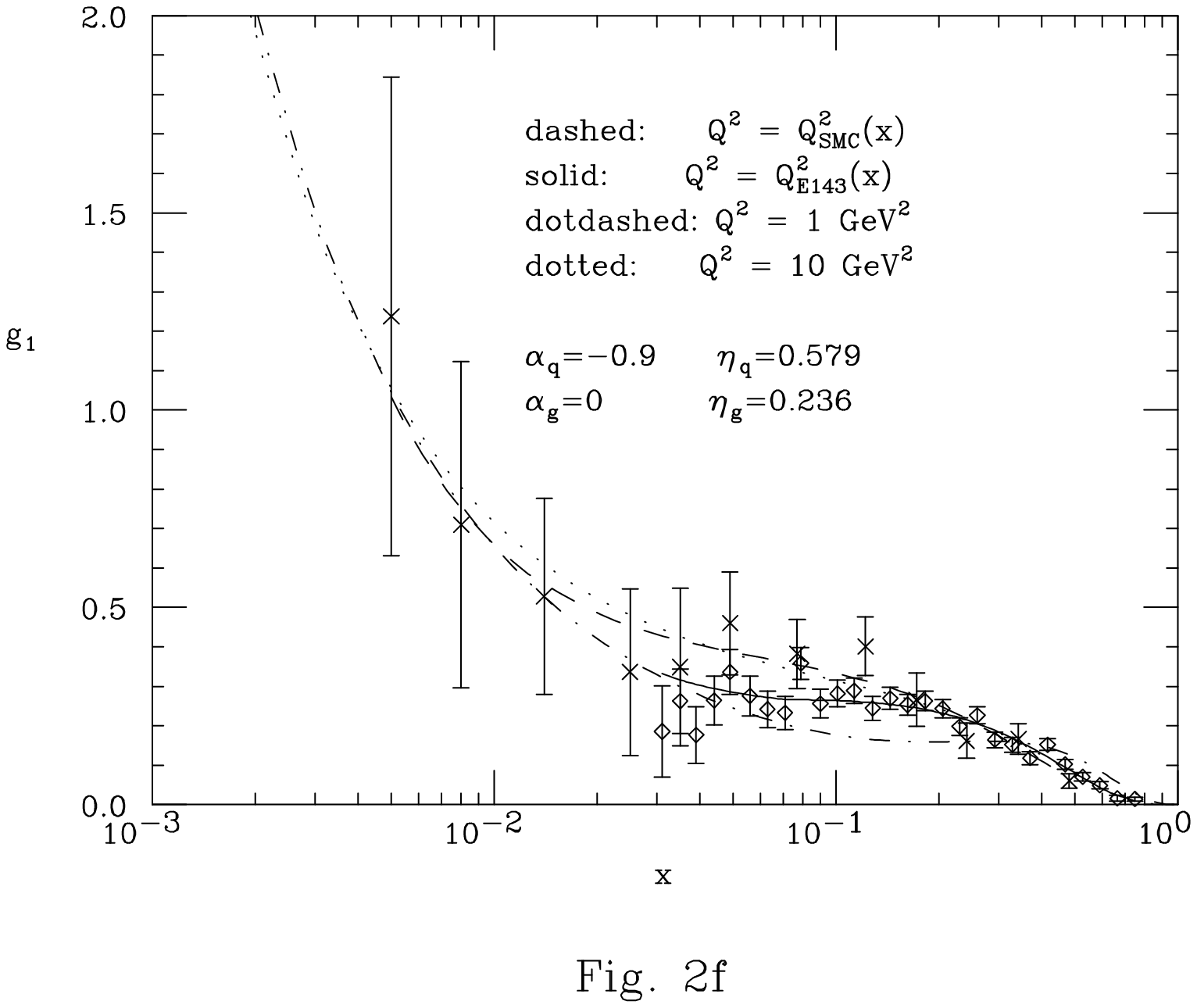}\hfil
\vfill
\eject
\null
\vskip -4.truecm
\epsfxsize=14truecm
\hfil\epsfbox{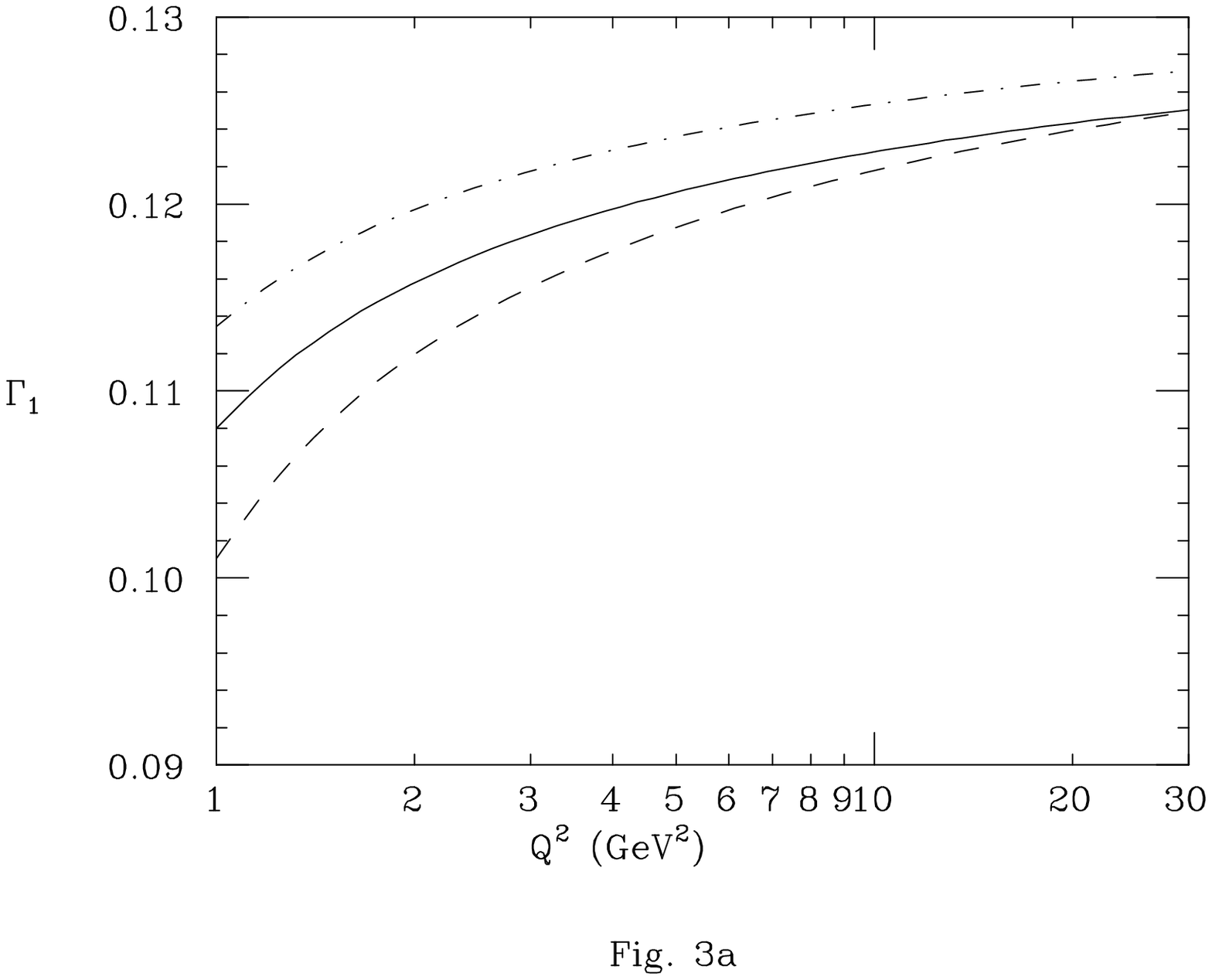}\hfil
\smallskip
\vskip -7.truecm
\epsfxsize=14truecm
\hfil\epsfbox{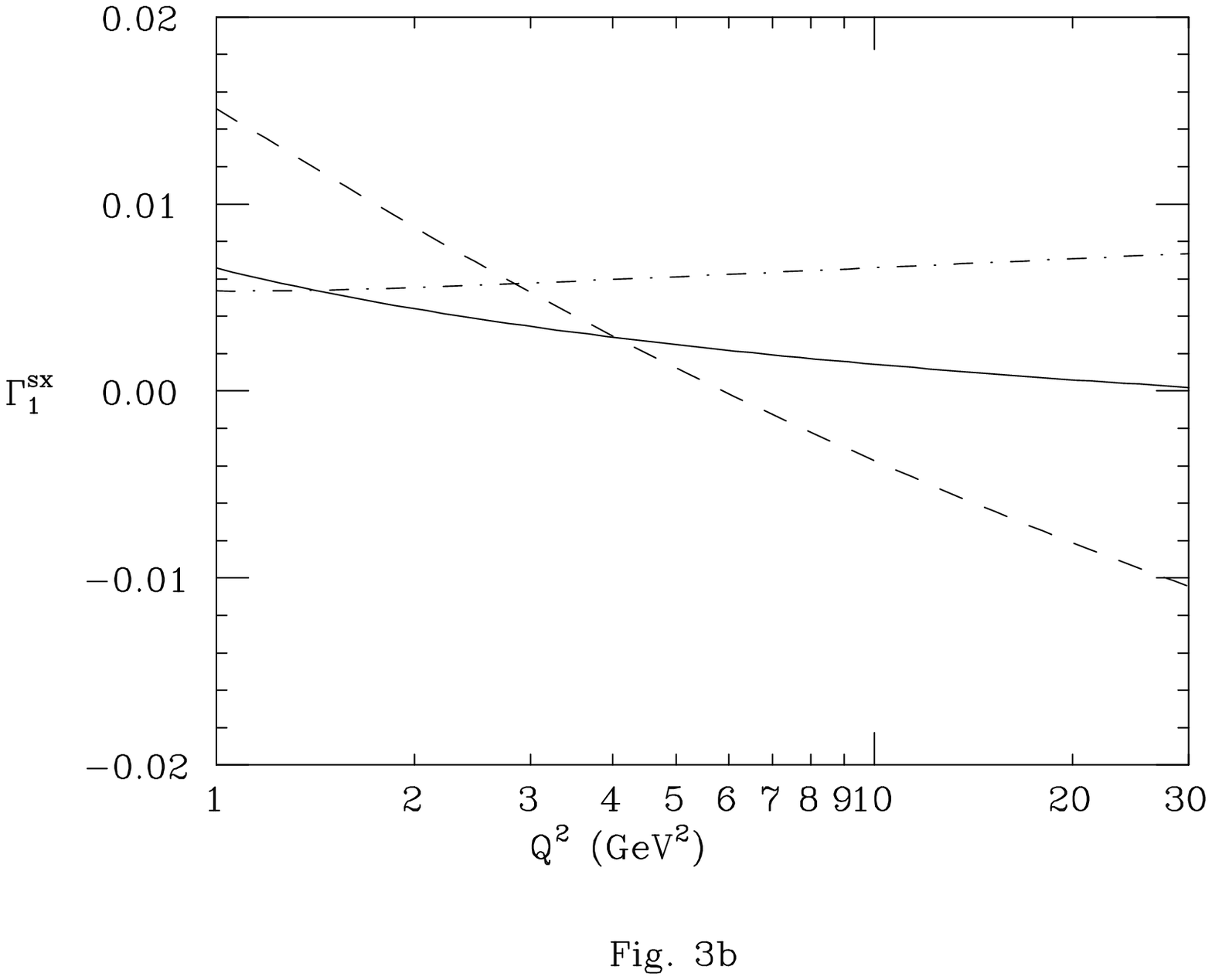}\hfil
\vfill
\eject
\bye